\documentclass[aps,pre,10pt,showpacs,floatfix,onecolumn,nofootinbib,a4paper]{revtex4-1}
\usepackage{amssymb, amsmath, amsthm, mathtools}
\usepackage{bm}
\usepackage{mathrsfs}
\usepackage{graphicx}
\usepackage{float}
\usepackage{caption}
\usepackage{subcaption}
\usepackage{verbatim}

\usepackage{tikz}

\newcommand{\R}{\mathbb{R}}
\renewcommand{\leq}{\leqslant}
\renewcommand{\geq}{\geqslant}

\newcommand{\tn}[1]{\mathbf{#1}}
\newcommand{\grad}{\nabla} 
\newcommand{\dv}{\operatorname{div}} 
\newcommand{\curl}{\operatorname{curl}} 
\newcommand{\tr}{\operatorname{tr}} 
\DeclareMathOperator{\tpr}{\otimes} 

\DeclareMathOperator{\sgn}{sgn}
\newcommand{\vn}{\bm{n}}
\newcommand{\n}{\bm{n}}
\newcommand{\vo}{\vn_{1}}
\newcommand{\vt}{\vn_{2}}
\newcommand{\basis}{(\vo,\vt,\vn)}
\newcommand{\vx}{\bm{e}_{x}}
\newcommand{\vy}{\bm{e}_{y}}
\newcommand{\vz}{\bm{e}_{z}}
\newcommand{\xyz}{(\vx,\vy,\vz)}
\newcommand{\bsplay}{\mathbf{D}}
\newcommand{\bend}{\bm{b}}
\newcommand{\dframe}{(\n_1,\n_2,\n)}
\newcommand{\gradn}{\grad\n}
\newcommand{\oct}{\mathbf{A}}
\newcommand{\x}{\bm{x}}
\newcommand{\sphere}{\mathbb{S}^2}

\begin{document}
	\title{Liquid Crystal Distortions Revealed by an Octupolar Tensor}
	\author{Andrea Pedrini}
	\email{andrea.pedrini@unipv.it}
	\author{Epifanio G. Virga}
	\email{eg.virga@unipv.it}
	\affiliation{Dipartimento di Matematica, Universit\`a di Pavia, Via Ferrata 5, 27100 Pavia, Italy}

	\date{\today}

	\begin{abstract}
The classical theory of liquid crystal elasticity as formulated by Oseen and Frank describes the (orientable) optic axis of these soft materials by a director $\n$. The ground state is attained when $\n$ is uniform in space; all other states, which have a non-vanishing gradient $\gradn$, are distorted. This paper proposes an algebraic  (and geometric) way to describe the local distortion of a liquid crystal by constructing from $\n$ and $\gradn$ a third-rank, symmetric and traceless tensor $\oct$ (the \emph{octupolar} tensor). The (nonlinear) eigenvectors of $\oct$ associated with the local maxima of its cubic form $\Phi$ on the unit sphere (its octupolar \emph{potential}) designate the directions of  \emph{distortion concentration}. The octupolar potential is illustrated geometrically and its symmetries are charted in the space of  \emph{distortion characteristics}, so as to educate the eye to capture the dominating elastic modes. Special distortions are studied, which have everywhere either the same octupolar potential or one with the same shape, but differently inflated.
	\end{abstract}
	
	\maketitle

\section{Introduction}\label{sec:intro}
An \emph{octupolar} tensor is a special third-rank tensor. The main objective of this paper is to justify the use of such a seemingly complicated tool to represent  elastic distortions in liquid crystals. Surely enough, higher-rank tensors are not new in condensed matter physics. For example,
the Landau approach in the theory of phase transitions is based on the identification of an \emph{order parameter} that
distinguishes the states of matter in the proximity of a critical point where a transition occurs. Although the order parameter can be either a scalar or a vector or, more generally, a tensor of any rank, it is common practice in the study of liquid crystals to select a second-rank tensor to represent the state of the medium when the constituent molecules resemble elongated rods. The fairly recent discovery of materials presenting a tetrahedral symmetry \cite{fel:tetrahedral,fel:symmetry} suggested the use of a \emph{fully symmetric} and \emph{completely traceless} third-order tensor, which we call \emph{octupolar} by analogy with electrostatics, to encode the variety of their possible phases \cite{liu:classification,liu:generalized,liu:generic}. 

However, this is not the only possible use of octupolar tensors. Besides reflecting molecular symmetries on the mesoscopic scale where the phase collective behavior is described, they may as well play a role irrespective of the symmetry of the molecular constituents. Here we illustrate a further application of octupolar tensors in soft matter physics building upon some earlier work \cite{virga:octupolar_2D,gaeta:octupolar,chen:octupolar,gaeta:symmetries}, which we shall often refer to, although our present approach will be  different to some extent.

In classical liquid crystal theory, the nematic director field $\vn$ describes the average orientation of the molecules that constitute the medium;  the elastic distortions of $\vn$ are locally measured by its gradient $\grad\vn$, which may become singular where the director exhibits \emph{defects} arising from a degradation of molecular order. The two main descriptors, $\vn$ and $\grad\vn$, can be combined into  the third-rank octupolar tensor
\begin{equation}\label{eq:octupolar_tensor}
\tn{A}:=\overbracket{\grad\vn\otimes\vn},
\end{equation}
where the superimposed hat $\overbracket{\cdots}$ makes  the tensor underneath it fully symmetric and traceless. It is perhaps worth noticing that $\oct$ is invariant under the change of orientation of $\n$, and so it duly enjoys the nematic symmetry, which makes it a good candidate for measuring intrinsically the local distortions of a director field.

Selinger \cite{selinger:interpretation}, extending earlier work \cite{machon:umbilic},
suggested a new interpretation of the elastic modes for nematic liquid crystals described by the Oseen-Frank elastic free energy, which penalizes in a quadratic fashion the distortions of $\n$ away from any uniform state. The Oseen-Frank energy-density is defined as (see, e.g., \cite[Ch.\,3]{deGennes:physics} and \cite[Ch.\,3]{virga:variational})
\begin{equation}\label{eq:frank_energy}
F := \frac{1}{2}K_{11}(\dv\vn)^{2} + \frac{1}{2}K_{22}(\vn\cdot\curl\vn)^{2} + \frac{1}{2}K_{33}|\vn\times\curl\vn|^{2} + K_{24}\big(\tr(\grad\vn)^{2}-(\dv\vn)^{2}\big),
\end{equation}
where $K_{11}$, $K_{22}$, $K_{33}$, and $K_{24}$ are the \emph{splay}, \emph{twist}, \emph{bend}, and \emph{saddle-splay} constants,  respectively, each associated with a corresponding elastic mode.\footnote{The saddle-splay term  is a null Lagrangian \cite{ericksen:nilpotent} and an integration over the bulk  reduces it to a surface energy. Here, however, the surface-like nature of $K_{24}$ will not be exploited.} 

The decomposition of $F$ in independent elastic modes proposed in \cite{selinger:interpretation} is achieved through a new decomposition of $\gradn$. If we denote by $\tn{P}(\vn)$ and $\tn{W}(\vn)$ the projection onto the plane orthogonal to $\vn$ and the skew-symmetric tensor with axial vector $\vn$, respectively, then 
\begin{equation}\label{eq:grad_n}
\grad\vn  = -\bend\otimes \vn + \frac{1}{2}T\tn{W}(\vn) + \frac{1}{2}S\tn{P}(\vn) + \bsplay,
\end{equation}
where $\bend := -(\grad\vn)\vn = \vn\times\curl\vn$ is the \emph{bend} vector, $T := \vn\cdot\curl\vn$ is the \emph{twist} (a pseudoscalar), $S := \dv\vn$ is the \emph{splay} (a scalar), and $\bsplay$ is a symmetric tensor such that $\bsplay\n=\bm{0}$ and $\tr\bsplay=0$.\footnote{For the only purpose of achieving a uniform notation, which does not mix alphabets, we shall denote this tensor with $\bsplay$, instead of the Greek counterpart $\bm{\Delta}$ used in both \cite{machon:umbilic} and \cite{selinger:interpretation}. The only other point where our notation differs from that of \cite{selinger:interpretation} is in calling $\bend$ the  bend vector, which there was $\bm{B}$.} The properties of $\bsplay$ guarantee that when $\bsplay\neq\bm{0}$ it can be represented as
\begin{equation}\label{eq:biaxial_splay_reoresentation}
\bsplay=q\left(\n_1\otimes\n_1-\n_2\otimes\n_2\right),
\end{equation}
where $q$ is the \emph{positive} eigenvalue of $\bsplay$. Following \cite{selinger:interpretation}, we shall call $q$ the \emph{biaxial splay}. The choice of sign for $q$ identifies (to within the orientation) the eigenvectors $\n_1$ and $\n_2$ of $\bsplay$ orthogonal to $\n$. Since $\tr\bsplay^2=2q^2$, we easily obtain from \eqref{eq:grad_n} that
\begin{equation}\label{eq:q_formula}
2q^2 = \tr(\grad\vn)^{2} + \frac{1}{2}T^{2} - \frac{1}{2}S^{2}.
\end{equation}
The Oseen-Frank  elastic free-energy density \eqref{eq:frank_energy} can be given the form
\begin{equation}\label{eq:frank_energy_selinger}
F = \frac{1}{2}(K_{11}-K_{24})S^{2} + \frac{1}{2}(K_{22}-K_{24})T^{2} + \frac{1}{2}K_{33}b^2 + K_{24}(2q^2),
\end{equation}
where all quadratic contributions are independent from one another.

The first advantage of such an expression is that it explicitly shows when the free energy  is positive semi-definite; thus is the case when the following inequalities, due to  Ericksen \cite{ericksen:inequalities}, are satisfied,
\begin{equation}\label{eq:ericksen}
K_{11} \geq K_{24} \geq 0,
\quad
K_{22} \geq K_{24} \geq 0,
\quad
K_{33} \geq 0.
\end{equation} 
Whenever $q>0$, the frame $\dframe$ is identified to within a change of sign in either $\n_1$ or $\n_2$; requiring that $\n=\n_1\times\n_2$, we reduce this ambiguity to a simultaneous change in the orientation of $\n_1$ and $\n_2$.\footnote{In this frame, $\tn{P}(\vn)=\tn{I}-\vn\tpr\vn$ and $\tn{W}(\vn)=\vt\tpr\vo-\vo\tpr\vt$.}  Since $\bend\cdot\n\equiv0$, we can represent $\bend$ as $\bend=b_1\n_1+b_2\n_2$. We shall call $\dframe$ the \emph{distortion frame} and $(S,T,b_1,b_2,q)$ the \emph{distortion characteristics} of the director field $\n$ \cite{virga:uniform}. In terms of these,  \eqref{eq:grad_n} can also be written as
\begin{equation}\label{eq:grad_in_basis}
\grad\vn  = \left(\frac{S}{2}+q\right)\vo\otimes\vo - \frac{T}{2}\vo\otimes\vt - b_{1}\vo\otimes\vn + \frac{T}{2}\vt\otimes\vo + \left(\frac{S}{2}-q\right)\vt\otimes\vt - b_{2}\vt\otimes\vn.
\end{equation}
Both \eqref{eq:grad_n} and \eqref{eq:grad_in_basis} show an intrinsic
decomposition of $\grad\vn$ into four genuine bulk contributions, namely, bend,  splay, twist, and biaxial splay.

The aim of this work is to explore the properties of the octupolar tensor in \eqref{eq:octupolar_tensor} and try and establish a qualitative, even visual way of representing through it the independent components of $\gradn$, hoping to educate the eye to recognize which components, if any, are predominant in a given distortion. Having symmetrized $\oct$, we have renounced to represent $T$, so no sign of twist will be revealed by our construction.\footnote{$T$ is a measure of chirality, and so it cannot be associated with a symmetric tensor. By  forming the completely skew-symmetric part of $\gradn\otimes\n$, one would  obtain the tensor $-\frac16T\bm{\varepsilon}$, where $\bm{\varepsilon}$ is Levi-Civita's alternator, the most general skew-symmetric, third-rank tensor in 3D.} 

By following \cite{gaeta:octupolar} and \cite{gaeta:symmetries}, in Sec.~\ref{sec:potential} we introduce the  \emph{octupolar potential}, a real-valued function defined on the unit sphere $\sphere$ which encodes all main properties of $\tn{A}$ and lends itself to a geometric $3\mathrm{D}$ representation. We single out the representations  where only one mode out of splay, bend, and biaxial splay is active. In Sec.~\ref{sec:symmetries}, we study how the elastic modes are related to the symmetry properties of the potential and to the number of its local maxima (and conjugated minima), which will be identified with the directions of local \emph{distortion concentration}. In Sec.~\ref{sec:defects} we discuss the octupolar potential and its geometric representation  for special director distortions. These include the \emph{uniform} distortions, for which all distortion characteristics are constant in space \cite{virga:uniform}, and new types of distortions, which we shall call \emph{quasi-uniform}. In the last Sec.~\ref{sec:conclusion} we draw the conclusions of our work. We leave detailed computations and proofs for a closing  Appendix~\ref{sec:appendix}. A good deal of our endeavour is graphical.

\section{Octupolar distortion potential}\label{sec:potential}
The \emph{octupolar potential} is the real-valued function defined as
\begin{equation}\label{eq:octupolar_potential}
\Phi(\bm{x}) := \tn{A}\cdot\bm{x}\otimes\bm{x}\otimes\bm{x} = \sum_{i,j,k=1}^{3}A_{ijk}x_{i}x_{j}x_{k},
\end{equation} 
where $\bm{x}=x_{1}\vo+x_{2}\vt+x_{3}\vn$ is a point on the unit sphere $\mathbb{S}^{2}\subset\R^{3}$ referred to the distortion frame $\dframe$, and $\tn{A}$ is the completely symmetric and traceless third-rank tensor defined in \eqref{eq:octupolar_tensor}. When we say that the
the octupolar potential $\Phi$ encodes the properties of the octupolar tensor, we mean  that the extremal points of $\Phi$ are precisely the eigenvectors of $\tn{A}$ on the sphere $\mathbb{S}^{2}$. Moreover, the value $\Phi(\bm{x})$ at an eigenvector $\bm{x}$ coincides with its associated eigenvalue $\lambda$ (see, for example,  \cite{gaeta:symmetries}).\footnote{For a complete collection of definitions and results concerning the  (nonlinear) eigenvalues and eigenvectors of a third-rank tensor, we refer the reader to the paper \cite{walcher:eigenvectors} and the textbook \cite{qi:tensor}, which also contains applications to liquid crystal theory.}

It is immediately seen that  $\Phi(\bm{x})=-\Phi(-\bm{x})$ for all $\bm{x}\in\mathbb{S}^{2}$, so that also the set of critical points of $\Phi$ is \emph{centrally symmetric}, with every local maximum $\bm{x}_{M}$ having a conjugate local minimum $\bm{x}_{m}=-\bm{x}_{M}$. Each local maximum $\bm{x}_{M}$ of $\Phi$ identifies the entire line $\gamma\bm{x}_{M}$ (with $\gamma\in\R$) of eigenvectors of $\tn{A}$ with eigenvalue $\lambda=\pm\Phi(\bm{x}_{M})$. We interpret the direction of $\x_M$ (and $\x_m$) as a direction of \emph{distortion concentration} for the nematic field $\vn$. Thus, we are  not interested in the sign of the eigenvalue $\lambda$, but only in its magnitude.\footnote{Similarly, one could argue that the octupolar potential itself is defined to within a sign, so that the  distinction between its maxima and minima would be artificial.} With this in mind, in Sec.~\ref{sec:symmetries}, we shall focus on the symmetry properties of $|\Phi|$ .

It was proved in \cite{gaeta:octupolar}  that the maximum number of non-degenerate critical points of $\Phi$ is $14$; in particular, the number of maxima can be either $3$ or $4$. In both \cite{gaeta:octupolar} and \cite{gaeta:symmetries}, the octupolar potential $\Phi$ was studied in a  frame that was judiciously chosen. In particular, the results on the cardinality of maxima (and minima) were obtained by \emph{orienting} the potential, which amounted to assume that one of its maxima (which exists, provided $\Phi\neq0$) falls at a prescribed point on the unit sphere $\sphere$, say the North Pole. Since the distortion frame $\dframe$ has here an intrinsic meaning, we no longer have the luxury of orienting $\Phi$. Indeed, a distinctive feature of our approach is to describe how the octupolar potential is \emph{oriented} relative to the distortion frame, so as to attribute a definite meaning to the directions of distortion concentration.

Here is how we shall represent graphically  the octupolar potential. Since $\Phi$ is centrally symmetric, in all the figures of this paper we depict the octupolar potential as the surface $\{\Phi(\bm{x})\bm{x}\ |\ \bm{x}\in\mathbb{S}^{2}\}$: each point $\bm{x}$ on the unit sphere is rescaled by the value of $\Phi$ at $\bm{x}$. In such \emph{polar plots}, we also draw  the frame $\dframe$, oriented so that $\n_1\times\n_2\cdot\n=+1$. Since no intrinsic length scale is associated with $\Phi$, to improve readability, we shall extend the arrows of the distortion frame so as to make them clearly discernible.

By use of \eqref{eq:grad_in_basis},
the octupolar potential becomes
\begin{equation}\label{eq:extended_potential}
 \Phi(\bm{x}) =\left(\frac{S}{2}+q\right)x_{1}^{2}x_{3} - b_{1} x_{1}x_{3}^{2} - b_{2} x_{2}x_{3}^{2} + \left(\frac{S}{2}-q\right)x_{2}^{2}x_{3} + \frac{1}{5}\big(x_{1}^{2}+x_{2}^{2}+x_{3}^{2}\big)\big(b_{1} x_{1}+b_{2} x_{2}-Sx_{3}\big).
\end{equation}
 As expected,  $\Phi$ does not depend on the twist $T$, but it does depend on the  biaxial splay $q$, which will play a prominent role in the following.
 
 It will be often more convenient to represent the bend vector $\bend$ (when it does not vanish) as $\bend=b\cos\beta\n_1+b\sin\beta\n_2$, with $b>0$ and $\beta\in]-\frac\pi2,\frac\pi2]$. This choice has two advantages: first, we can write
 \begin{equation}\label{eq:bend_components}
 b_1=b\cos\beta,\quad b_2=b\sin\beta;
 \end{equation}
 second, the orientation of $\n_2$ (and so that of $\n_1$) is prescribed, as $\n_1\cdot\bend\geq0$.\footnote{Occasionally, to enhance the graphical symmetry of some figures, we shall also allow $\beta\in]\frac\pi2,\frac{3\pi}{2}]$. In such cases, for consistency, both $\n_1$ and $\n_2$ should be meant as being simultaneously reversed.}
Sometimes, it is useful  to stress the dependency of $\Phi$ on the distortion characteristics; in these cases, we shall use the extended notation $\Phi(\bm{x};S,b,\beta,q)$, where use  of \eqref{eq:bend_components} is to be understood.

\subsection{Pure Modes}\label{sec:pure_modes}
Here we describe  the octupolar potential $\Phi$ in the very special cases where one and only one elastic mode is exhibited. 
In particular, we explore two different but related geometrical properties: the symmetries enjoyed by the polar plots of  $\Phi$ and the number of its maxima. The following discussion shall be made more formal and more general in the subsequent sections.

\begin{figure}[h]
\centering
 \begin{subfigure}[b]{0.3\textwidth}
 \centering
  \includegraphics[width=\textwidth]{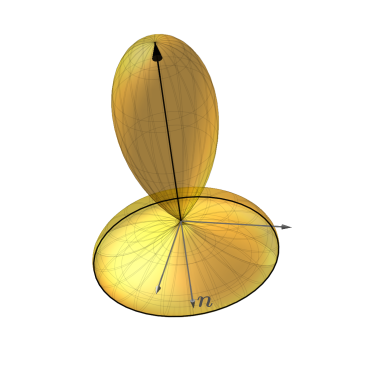}
  \caption{Pure splay}
  \label{fig:octu_s}
 \end{subfigure}
 $\quad$
 \begin{subfigure}[b]{0.3\textwidth}
 \centering
  \includegraphics[width=\textwidth]{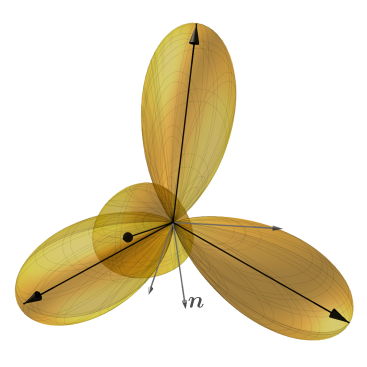}
  \caption{Pure biaxial-splay}
  \label{fig:octu_d}
 \end{subfigure}
 $\quad$
 \begin{subfigure}[b]{0.3\textwidth}
 \centering
  \includegraphics[width=\textwidth]{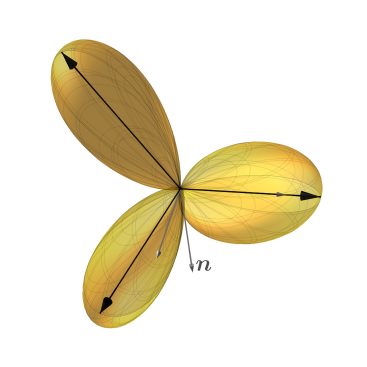}
  \caption{Pure bend}
  \label{fig:octu_b}
 \end{subfigure}
\caption{Polar plots of the octupolar potential for pure elastic modes. The long, black arrows represent the maxima.}
 \label{fig:pure_modes}
\end{figure}

\subsubsection{Splay}
When splay is the only active mode, the choice of $\vo$ and $\vt$ in the plane orthogonal to $\vn$ is arbitrary. This fact reverberates in the symmetries of the octupolar potential and also in its critical points. In this case,
\begin{equation}
 \Phi(\bm{x}) = \frac{1}{10}S\big(3x_{1}^{2}x_{3} + 3x_{2}^{2}x_{3} - 2x_{3}^{3}\big).
\end{equation}
Graphically, $\Phi(\bm{x})$ is depicted in Fig.~\ref{fig:octu_s}; it has a big lobe elongated around $\vn$ and a perpendicular circular pedestal. More precisely, the apex of the lobe is at $\bm{x}=-\vn$ (when $S$ is positive) and coincides with the absolute maximum of the potential, with value $\frac{1}{5}S$, while the pedestal is the ring $\bm{x}=x_{1}\vo + x_{2}\vt + \frac{1}{\sqrt{5}}\vn$ (with $x_{1}^{2}+x_{2}^{2}=\frac{4}{5}$) of equal local maxima with value $\frac{1}{5\sqrt{5}}S$.
It is easy to check that such a potential is invariant under all rotations around $\vn$ and all reflections with mirror plane containing $\vn$. 

\subsubsection{Biaxial splay}
When both $S=0$ and $b=0$, but $q>0$, the potential is
\begin{equation}
 \Phi(\bm{x}) = q(x_{1}^{2}-x_{2}^{2})x_{3}.
\end{equation}
Figure~\ref{fig:octu_d} shows that $\Phi(\bm{x})$ has four identical lobes, spatially distributed as the vertices of a regular tetrahedron. Accordingly, its maxima are the four points $\frac{1}{\sqrt{3}}(\pm\sqrt{2}\vo + \vn)$ and $\frac{1}{\sqrt{3}}(\pm\sqrt{2}\vt - \vn)$, each with  value $\frac{2q}{3\sqrt{3}}$.

\subsubsection{Bend}
For pure bend, we can choose $\vo$ and $\vt$ such that $\bend=b\vo$ with $b>0$. Then the potential,
\begin{equation}
 \Phi(\bm{x}) = \frac{1}{5}bx_{1}\left(x_{1}^{2}+x_{2}^{2}-4x_{3}^{2}\right) = \frac{b}{5}x_{1}\left(1-5x_{3}^{2}\right),
\end{equation}
has three lobes: two larger, with equal  height $\frac{16b}{15\sqrt{15}}$ at $\bm{x}=\frac{1}{\sqrt{15}}\left(-2\vo\pm\sqrt{11}\vn\right)$, and one smaller at $\bm{x}=\vo$ with height $\frac{b}{5}$. As shown in Fig.~\ref{fig:octu_b}, the polar plot of $\Phi$ is invariant under both a rotation by angle $\pi$ around $\vo$ and the mirror symmetry with respect to the plane containing $(\n_1,\n)$.

\begin{figure}[h]
	\centering
	\includegraphics[width=0.3\textwidth]{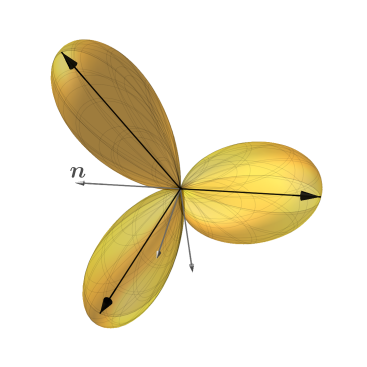}
	\caption{Polar plot of the octupolar potential for a combination of splay and biaxial splay ($S=2q$) which reproduces the pure bend case, except that for the  orientation in the distortion frame $\dframe$.}
	\label{fig:bend_mimicked}
\end{figure}

Despite the attractive features of the octupolar potential when only one mode is involved, the reader should not be misled by Fig.~\ref{fig:pure_modes} to think that the pure geometry of the octupolar potential would suffice to reveal the prevalence of one mode with respect to the others in a general distortion.
Figure~\ref{fig:bend_mimicked}, for example, shows that the polar plot of $\Phi$ in case of pure bend can be perfectly mimicked by a combination of splay and biaxial splay.
As is easily  checked, for all $\bm{x}\in\mathbb{S}^{2}$,
\begin{equation}
 \Phi\left(\bm{x};b,0,-,b/2\right) = -\frac{1}{5}bx_{3}\left(1-5x_{1}^{2}\right) = \Phi(\bm{y};0,b,0,0),
\end{equation} 
where $\bm{y}$ is   $\bm{x}$ rotated by $\frac{\pi}{2}$ around $\vt$.\footnote{Here (and below) with a ``$-$'' we mean that the corresponding argument of $\Phi$ can be chosen arbitrarily.}
To distinguish the two cases, it is crucial to know how  $\Phi$ is oriented in the  distortion frame $\dframe$: for $S=b=2q$, the $\pi$-rotation symmetry axis is indeed  $\vn$, and not $\bend$, as for the  case of pure bend. This seemingly unimportant difference has further  consequences, as we shall see in Sec.~\ref{sec:defects}.

\section{Symmetries}\label{sec:symmetries}
To explore the significance of the octupolar potential $\Phi$ and our preferred  graphical representation (the polar plot) in illustrating all distortion characteristics but $T$, we study here the symmetries enjoyed by $\Phi$. Actually, as anticipated,we are interested in the symmetries of $|\Phi|$. Our analysis will also echo that in \cite{gaeta:symmetries}, but it shall also differ from that, mainly because the distortion frame is here intrinsically prescribed.

We give a formal description of the groups representing the symmetries enjoyed by the octupolar potential when considering suitable combinations of elastic modes. For this entire section, $\bm{e}$ and $\bm{e}_{\perp}$ shall be two orthogonal unit vectors in $\R^{3}$. We denote by $\tn{Q}_{\delta,\bm{e}}$ the rotation by angle $\delta$ around the axis identified by $\bm{e}$ and by $\tn{R}_{\bm{e}}$ the reflection with respect to the plane orthogonal to $\bm{e}$. Formally,
\begin{equation}
\begin{split}
 \tn{Q}_{\delta,\bm{e}} &:= \tn{I} + \sin\delta\tn{W}(\bm{e}) - (1-\cos\delta)\tn{P}(\bm{e}),\\
 \tn{R}_{\bm{e}}&:=\tn{I}-2\bm{e}\otimes\bm{e},
\end{split}
\end{equation}
where $\tn{I}$ is the identity tensor.

We shall say that the octupolar potential is \emph{invariant} under the \emph{symmetry} $\tn{Q}_{\delta,\bm{e}}$ or $\tn{R}_{\bm{e}}$ if $|\Phi(\tn{Q}_{\delta,\bm{e}}\bm{x})|=|\Phi(\bm{x})|$ or $|\Phi(\tn{R}_{\bm{e}}\bm{x})|=|\Phi(\bm{x})|$, respectively, for all $\bm{x}\in\mathbb{S}^{2}$. Such a definition ensures that the extrema of $\Phi$ are invariant under a symmetry, as are the directions of distortion concentration (i.e., the eigenvectors of $\tn{A}$).

\begin{figure}[h]
	\centering
	\includegraphics[width=0.7\textwidth]{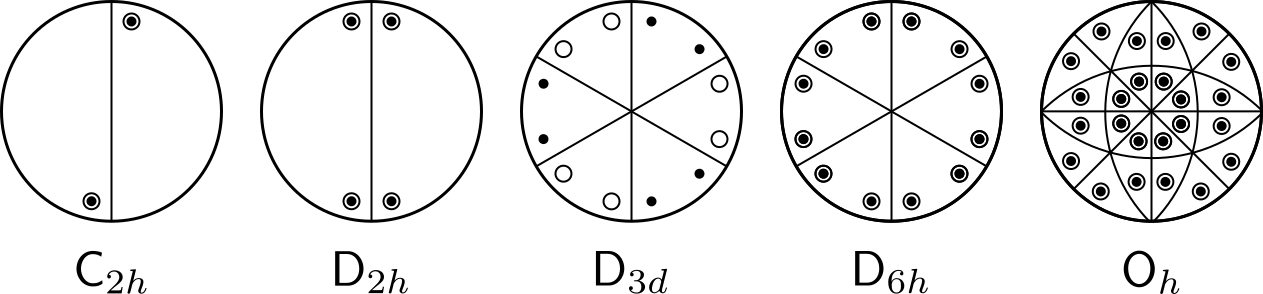}
	\caption{Stereograms representing the five finite point groups for the octupolar potential.}
	\label{fig:stereograms}
\end{figure}

The following six (point) \emph{symmetry groups} are relevant here;  in the Sch\"onflies notation (see, for example, \cite[Sec.\,3.10]{ladd:symmetry} and \cite[Sec.\,2.9]{hamermesh:group}), they read as
\begin{enumerate}
 \item $\mathsf{C}_{2 h}$, generated by $\tn{Q}_{\pi,\bm{e}}$ and $\tn{R}_{\bm{e}}$;
 \item $\mathsf{D}_{2 h}$, generated by $\tn{Q}_{\pi,\bm{e}}$, $\tn{R}_{\bm{e}}$ and $\tn{R}_{\bm{e}_{\perp}}$;
 \item $\mathsf{D}_{3d}$, generated by $\tn{Q}_{2\pi/3,\bm{e}}$, $\tn{Q}_{\pi,\bm{e}_{\perp}}$ and $\tn{R}_{\bm{e}_{\perp}}$;
 \item $\mathsf{D}_{6h}$, generated by $\tn{Q}_{\pi/3,\bm{e}}$, $\tn{R}_{\bm{e}}$ and $\tn{R}_{\bm{e}_{\perp}}$;
  \item $\mathsf{O}_{h}$, the symmetry group of the (regular)  octahedron (and the cube);
 \item $\mathsf{D}_{\infty h}$, generated by $\tn{Q}_{\delta,\bm{e}}$ for all $\delta\in[0,2\pi]$, $\tn{R}_{\bm{e}}$ and $\tn{R}_{\bm{e}_{\perp}}$.
\end{enumerate}
The schematic representation for the first five (finite) groups is given by the stereograms in Fig.~\ref{fig:stereograms}.
Each group contains the symmetries under which the corresponding diagram, considered as an object in the three dimensional space, is invariant. A black dot represents  a small bump protruding from the front of the page, while an empty circle represents an identical bump on the back of the page, so that the group contains the reflection with respect to the plane of the page if and only if dots and circles are paired, one inside the other. In this case, the name of the group has an $h$ as a subscript to indicate the presence of a \emph{horizontal} plane of reflection. The main axis $\bm{e}$ of rotation is orthogonal to the page, while the subscript numbers indicate the angle of rotation: $2$ for a $2$-folded rotation by $\pi$, $3$ for a $3$-folded rotation by $\frac{2\pi}{3}$, and so on. The vertical and horizontal pairs of arcs in the stereogram for $\mathsf{O}_{h}$ are to be intended as two circles orthogonal to each other and to the one on the plane of the page: all these three circle are of type $\mathsf{D}_{4h}$, which is similar to $\mathsf{D}_{6h}$, but with four sectors instead of six.

\begin{figure}[h]
	\centering
	\begin{subfigure}[b]{0.23\textwidth}
		\centering
		\includegraphics[width=\textwidth]{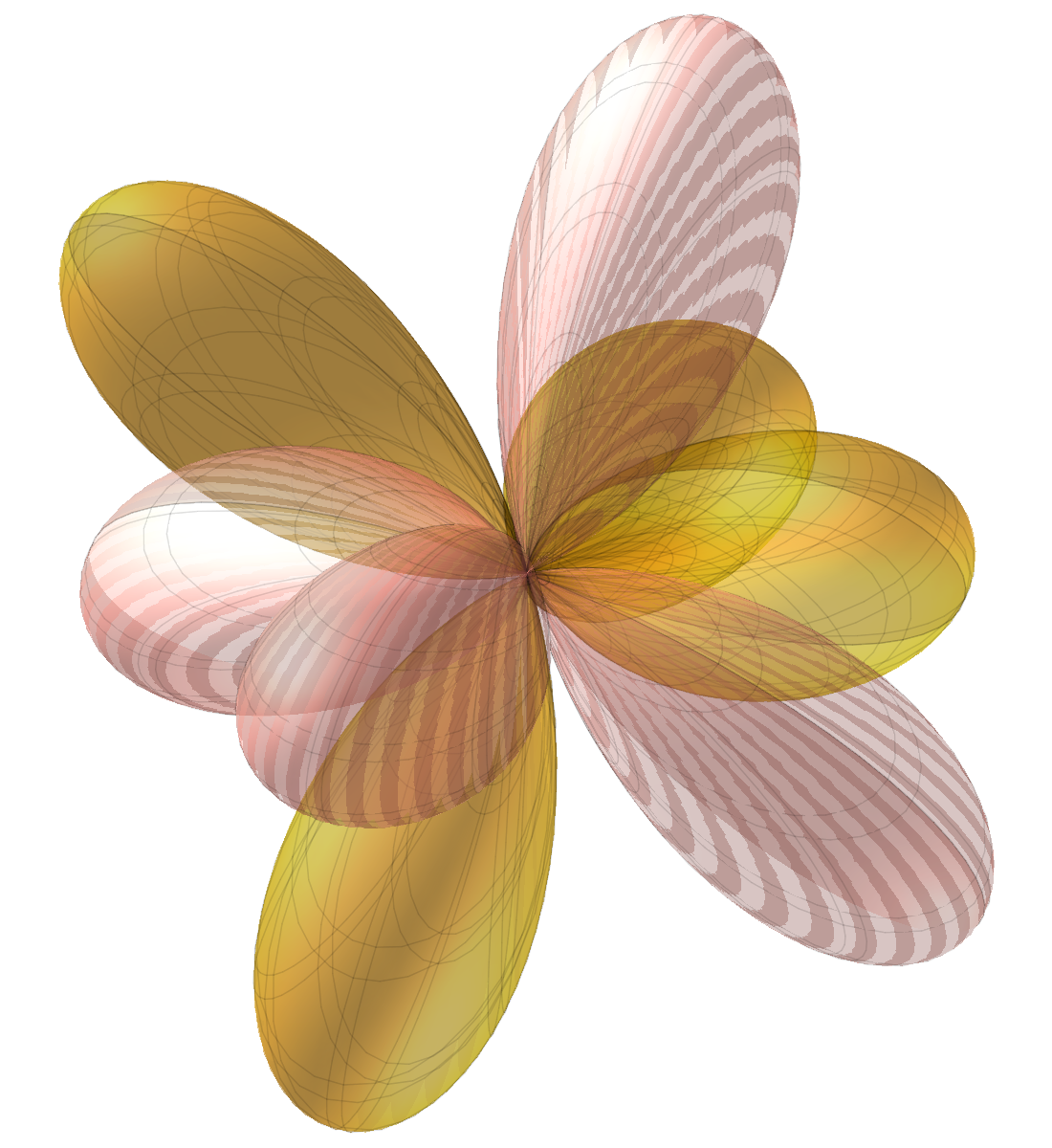}
		\caption{$\mathsf{O}_{h}$}
	\end{subfigure}
	$\qquad\qquad$
	\begin{subfigure}[b]{0.23\textwidth}
		\centering
		\includegraphics[width=\textwidth]{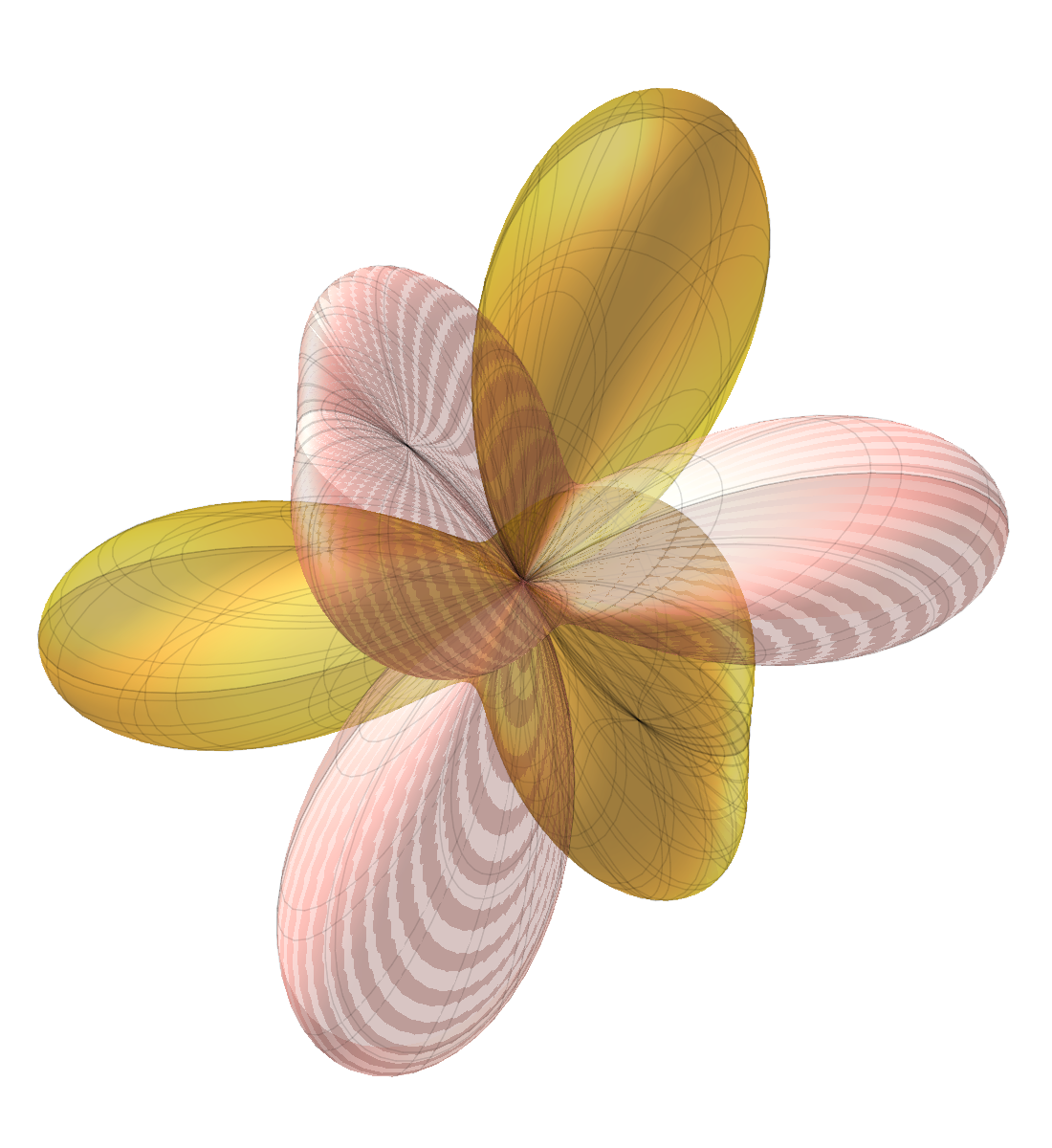}
		\caption{$\mathsf{C}_{2 h}$}
	\end{subfigure}
	$\qquad\qquad$
	\begin{subfigure}[b]{0.23\textwidth}
		\centering
		\includegraphics[width=\textwidth]{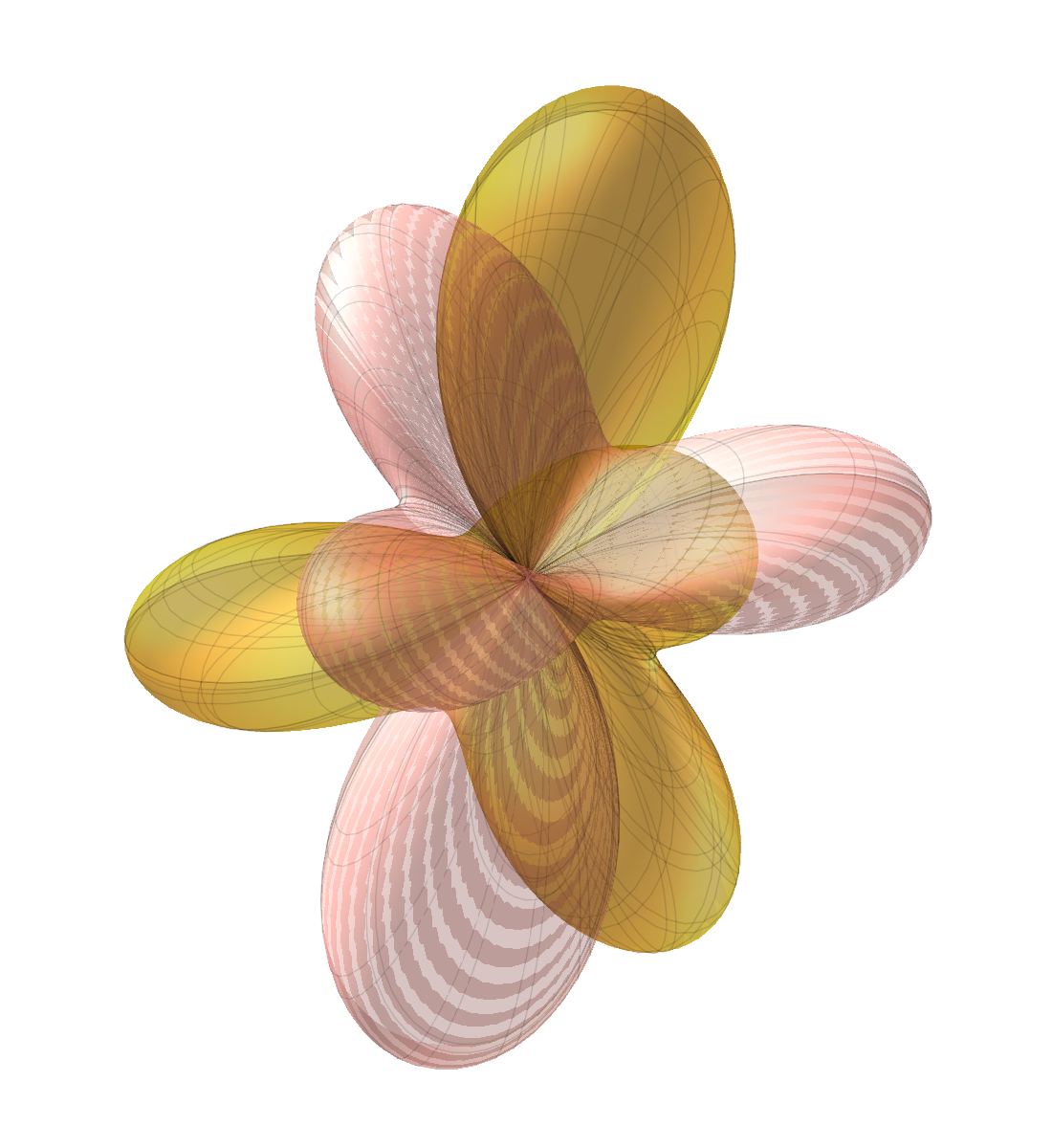}
		\caption{$\mathsf{D}_{3d}$}
		\label{fig:c3v_d}
	\end{subfigure}
	\caption{Combined polar plots of $\Phi$ (yellow) and $-\Phi$ (pink) for three cases of symmetry invariance.}
	\label{fig:double}
\end{figure}

A glance at Fig.~\ref{fig:pure_modes} suggests that each pure mode enjoys the invariance under one of the aforementioned symmetry group: $\mathsf{D}_{\infty h}$ for splay, $\mathsf{O}_{h}$ for biaxial splay, and $\mathsf{D}_{2 h}$ for bend. The reader should not be misled here by the polar plots of $\Phi$, in which minima are invaginated under maxima by construction, and keep in mind  that we are interested in the maxima of $|\Phi|$. To make this point clearer, Fig.~\ref{fig:double} shows three examples of symmetry invariance by plotting both $\Phi$ (in yellow) and $-\Phi$ (in pink): the union of the two surfaces is the right object to have in mind when looking for point symmetry groups.
However, such \emph{combined} polar plots are unnecessarily intricate and hereafter we shall stick to the choice of representing only the polar plot of $\Phi$, but with the implicit meaning expounded in Fig.~\ref{fig:double}. Accordingly, when we shall say that $\Phi$ is invariant under a certain group, we shall actually mean that the associated invariant geometric object is the corresponding polar plot as in Fig.~\ref{fig:double}.\footnote{In this respect, our analysis here sharpens that in \cite{gaeta:symmetries}. For a closer comparison, we should replace the tetrahedral group $\mathsf{T}_d$ used there with $\mathsf{O}_h$.}

In the general case, in the space of distortion characteristics we shall detect further loci with symmetry invariance of the octupolar potential. Before  describing them (in two separate cases, $q=0$ and $q\neq0$), we find it useful to list a set of properties involving octupolar potentials with different but related elastic modes. They are very easy to check (see Appendix~\ref{sec:appendix}) and  explain directly the peculiar interconnection between the trajectories with one and the same symmetry invariance that will be presented in Fig.~\ref{fig:symmetries} below. In our notation,
\begin{subequations}
\label{eq:symmetries}
 \begin{align}
  &\Phi(\bm{x};S,b,-\beta,q)=\Phi(\tn{R}_{\vt}\bm{x};S,b,\beta,q), \label{eq:sym1} \\
  &\Phi(\bm{x};-S,b,\pi/2-\beta,q,\bm{x})=-\Phi(\tn{R}_{\vt}\tn{Q}_{\frac{\pi}{2},\vn}\bm{x};S,b,\beta,q), \label{eq:sym2} \\
  &\Phi(\bm{x};S,b,\pi+\beta,q,\bm{x})=-\Phi(\tn{R}_{\vn}\bm{x};S,b,\beta,q). \label{eq:sym3}
 \end{align}
\end{subequations}
Properties \eqref{eq:symmetries} will also be useful in the next section, where we discus the number of critical points of $\Phi$.

\subsection{Vanishing  biaxial splay ($q=0$)}\label{sec:vanishing}
We describe here the octupolar potential $\Phi$ when $q=0$. In this case, $\vo$ and $\vt$ are no longer defined intrinsically in terms of $\grad\vn$, and so  we are at liberty of taking $\bend=b\vo$: any other choice of $\basis$ would  simply result in a rotation of the potential around $\vn$. Moreover, we can always rescale $\Phi$ and set $S=1$, thus obtaining a potential depending only on $b$,
\begin{equation}\label{eq:null_q}
 \Phi(\bm{x})=\frac{1}{2}x_{1}^{2}x_{3} - b x_{1}x_{3}^{2} + \frac{1}{2}x_{2}^{2}x_{3} + \frac{1}{5}\left(x_{1}^{2}+x_{2}^{2}+x_{3}^{2}\right)\left(b x_{1}-x_{3}\right).
\end{equation}
Except for the extreme cases $b=0$ (pure splay) and $b=\infty$ (pure bend), such a potential enjoys only the invariance under the symmetry group $\mathsf{C}_{2 h}$; it has \emph{three} maxima and four saddle points. The former property trivially derives from equation \eqref{eq:sym1}, as  $\Phi(\bm{x})=\Phi(\tn{R}_{\vt}\bm{x})$, and from a direct inspection of \eqref{eq:null_q}, as $\Phi(\bm{x})=-\Phi(\tn{Q}_{\pi,\vt}\bm{x})$.

\begin{figure}[h]
	\centering
	\begin{subfigure}[b]{0.23\textwidth}
		\centering
		\includegraphics[width=\textwidth]{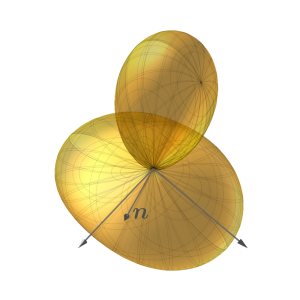}
		\caption{$b=0.3$}
		\label{fig:b03}
	\end{subfigure}
	$\ $
	\begin{subfigure}[b]{0.23\textwidth}
		\centering
		\includegraphics[width=\textwidth]{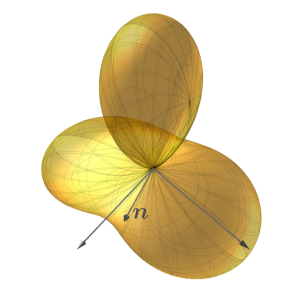}
		\caption{$b=0.7$}
		\label{fig:b07}
	\end{subfigure}
	$\ $
	\begin{subfigure}[b]{0.23\textwidth}
		\centering
		\includegraphics[width=\textwidth]{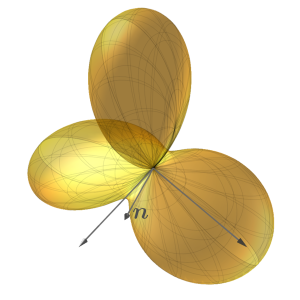}
		\caption{$b=1.5$}
		\label{fig:b15}
	\end{subfigure}
	$\ $
	\begin{subfigure}[b]{0.23\textwidth}
		\centering
		\includegraphics[width=\textwidth]{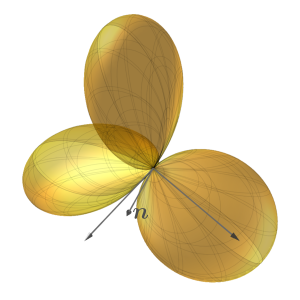}
		\caption{$b=5.0$}
		\label{fig:b5}
	\end{subfigure}
	\caption{Sequence  of polar plots  of $\Phi$ for $q=0$, $S=1$, and increasing values of  $b$. For all $b>0$ the potential has three maxima and is invariant under $\mathsf{C}_{2 h}$.}
	\label{fig:q_zero}
\end{figure}

Figure~\ref{fig:q_zero} shows the polar plots of \eqref{eq:null_q} for increasing $b$. 
It is manifest how a single lobe (on a roundish pedestal) oriented along $\n$ is soon converted into a three-lobe object protruding in a direction orthogonal to $\n$ (that of $\bend$).

We finally note that \eqref{eq:null_q} is more precisely $\Phi(\bm{x};1,b,0,0)$ and this  can also be seen as the limiting case for $|S|\to\infty$ or $b\to\infty$ (or both) of a generic $\Phi$. In other words, the case $q=0$  is essentially the special situation where splay and bend are much larger than biaxial splay.

\subsection{Non-vanishing biaxial splay ($q=1$)}
We consider now the more general and challenging case where biaxial splay is not zero. By rescaling to $q>0$ all other distortion characteristics, we may effectively set  $q=1$, with no loss of generality.

The best way to extract a number of distinctive features possessed in this case by the octupolar potential   is to go ``symmetry hunting''. We shall see that the landscape of distortion characteristics $(S,b,\beta)$  showing the same symmetries of $\Phi$ (thus having resemblant polar plots) is rather rich and not without aesthetic appeal.  

\begin{figure}[h]
	\centering
	\begin{subfigure}[b]{0.23\textwidth}
		\centering
		\includegraphics[width=\textwidth]{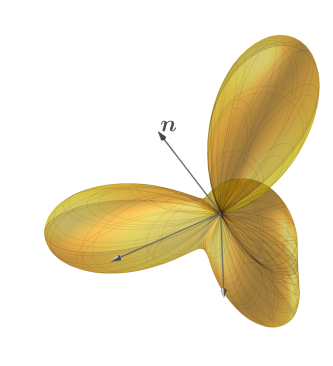}
		\caption{$\mathsf{C}_{2 h}$ symmetry, with\\$S=1$, $b=1$ and $\beta=0$}
		\label{fig:cs}
	\end{subfigure}
	$\ $
	\begin{subfigure}[b]{0.23\textwidth}
		\centering
		\includegraphics[width=\textwidth]{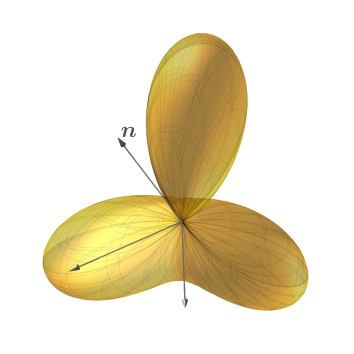}
		\caption{$\mathsf{D}_{2 h}$ symmetry, with\\$S=0$, $b=5/\sqrt{2}$ and $\beta=0$}
		\label{fig:c2v}
	\end{subfigure}
	$\ $
	\begin{subfigure}[b]{0.23\textwidth}
		\centering
		\includegraphics[width=\textwidth]{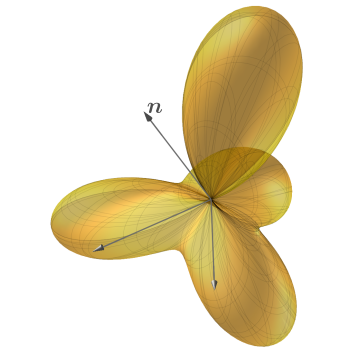}
		\caption{$\mathsf{D}_{3d}$ symmetry, with\\$S=0$, $b=\sqrt{5}/2$ and $\beta=0$}
		\label{fig:c3v}
	\end{subfigure}
	$\ $
	\begin{subfigure}[b]{0.23\textwidth}
		\centering
		\includegraphics[width=\textwidth]{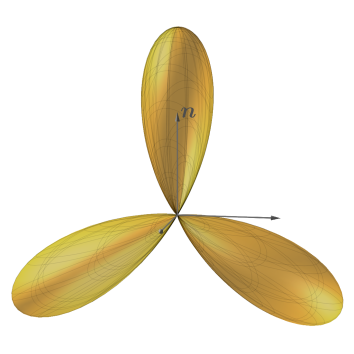}
		\caption{$\mathsf{D}_{6h}$ symmetry (monkey saddle), with\\$S=-10/3$ and $b=0$}
		\label{fig:monkey_saddle}
	\end{subfigure}
	\caption{Examples of symmetries enjoyed by the octupolar potential for $q=1$.}
	\label{fig:special_symmetries}
\end{figure}

\begin{figure}[h]
	\centering
	\begin{subfigure}[b]{0.27\textwidth}
		\centering
		\includegraphics[width=\textwidth]{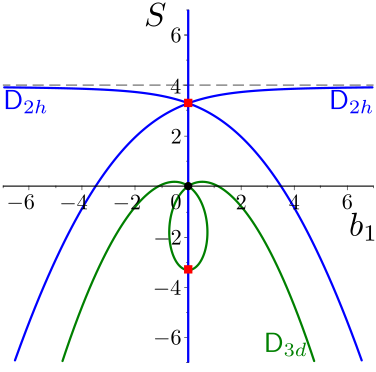}
		\caption{Plane $b_{2}=0$ ($\beta=0$).}
		\label{fig:sym_sb1}
	\end{subfigure}
	$\qquad$
	\begin{subfigure}[b]{0.27\textwidth}
		\centering
		\includegraphics[width=\textwidth]{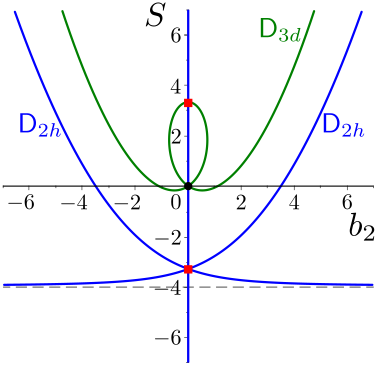}
		\caption{Plane $b_{1}=0$ ($\beta=\pi/2$).}
		\label{fig:sym_sb2}
	\end{subfigure}
	$\qquad$
	\begin{subfigure}[b]{0.27\textwidth}
		\centering
		\includegraphics[width=\textwidth]{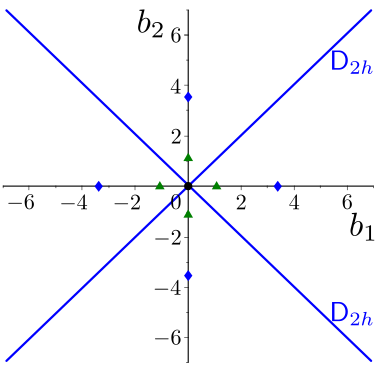}
		\caption{Plane $S=0$.}
		\label{fig:sym_b1b2}
	\end{subfigure}
	
	\vspace{0.5cm}
	
	\begin{subfigure}[b]{0.27\textwidth}
		\centering
		\includegraphics[width=\textwidth]{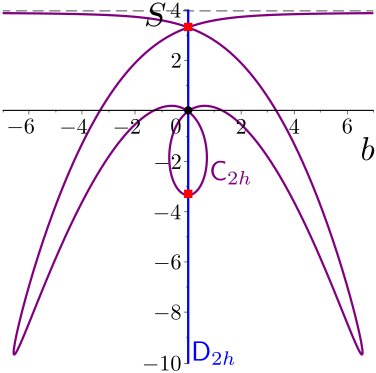}
		\caption{Plane $\beta=\pi/32$.}
		\label{fig:sym_Pi32}
	\end{subfigure}
	$\qquad$
	\begin{subfigure}[b]{0.27\textwidth}
		\centering
		\includegraphics[width=\textwidth]{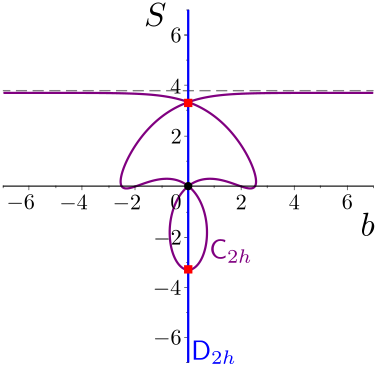}
		\caption{Plane $\beta=\pi/16$.}
		\label{fig:sym_Pi16}
	\end{subfigure}
	$\qquad$
	\begin{subfigure}[b]{0.27\textwidth}
	\centering
	\includegraphics[width=\textwidth]{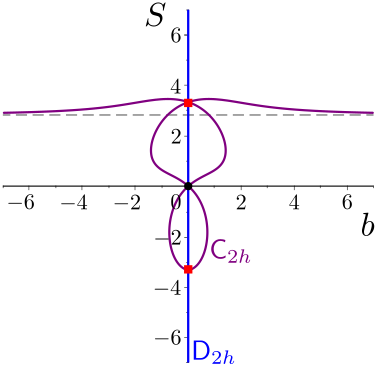}
	\caption{Plane $\beta=\pi/8$.}
	\label{fig:sym_Pi8}
\end{subfigure}
	
	\vspace{0.5cm}

\begin{subfigure}[b]{0.27\textwidth}
		\centering
		\includegraphics[width=\textwidth]{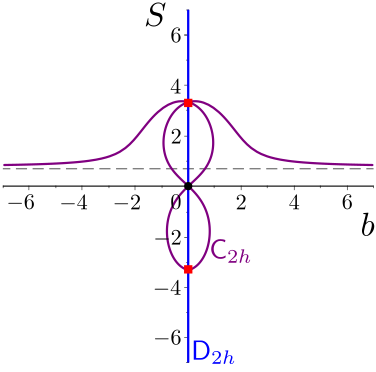}
		\caption{Plane $\beta=7\pi/32$.}
		\label{fig:sym_7Pi32}
	\end{subfigure}
	$\qquad$
	\begin{subfigure}[b]{0.27\textwidth}
		\centering
		\includegraphics[width=\textwidth]{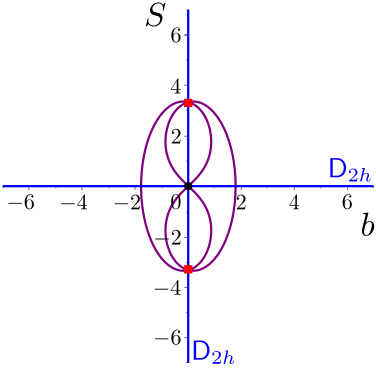}
		\caption{Plane $\beta=\pi/4$.}
		\label{fig:sym_Pi4}
	\end{subfigure}
$\qquad$
	\begin{subfigure}[b]{0.27\textwidth}
	\centering
	\includegraphics[width=\textwidth]{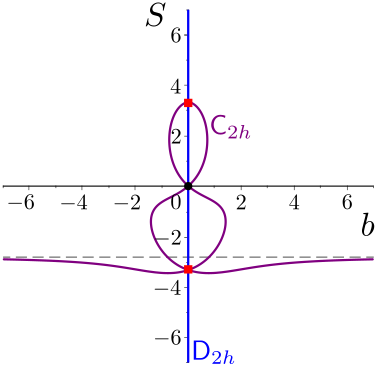}
	\caption{Plane $\beta=3\pi/8$.}
	\label{fig:sym_3Pi32}
\end{subfigure} 
	\caption{Trajectories of symmetry in the space $q=1$. The blue lines are for the $\mathsf{D}_{2 h}$ symmetry, while the green curves are for $\mathsf{D}_{3d}$ and the purple ones for $\mathsf{C}_{2 h}$. The monkey saddles with $\mathsf{D}_{6h}$ symmetry are the red squares;  the $\mathsf{O}_{h}$ symmetry  is enjoyed at the black dot only, corresponding to pure biaxial splay. The blue diamonds in panel (c) are the traces left on the $(b_1,b_2)$ plane by the $\mathsf{D}_{2h}$ lines in both panels (a) and (b); the green triangles in panel (c) are the traces left on the $(b_1,b_2)$ plane by the $\mathsf{D}_{3h}$ lines in both panels (a) and (b). All points in both planes $b_{2}=0$ and $b_{1}=0$ are invariant under $\mathsf{C}_{2 h}$. The marked values of $\beta$ correspond to $b\geq0$. For $b\leq0$, all these graphs represent the symmetric branches corresponding to $\beta+\pi$.}
	\label{fig:symmetries}
\end{figure}

\begin{figure}[h]
	\centering
	\begin{subfigure}[b]{0.23\textwidth}
		\centering
		\includegraphics[width=\textwidth]{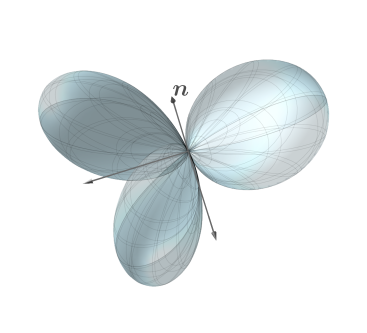}
		\caption{$S=3.97$ and $b=-7.68$}
		\label{fig:c2v_progr01}
	\end{subfigure}
	$\qquad\qquad$
	\begin{subfigure}[b]{0.23\textwidth}
		\centering
		\includegraphics[width=\textwidth]{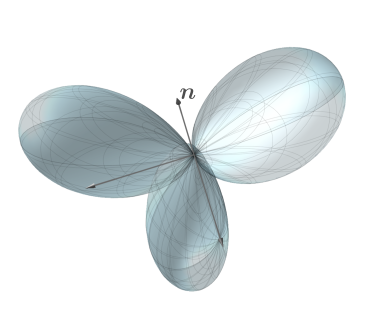}
		\caption{$S=3.78$ and $b=-1.98$}
		\label{fig:c2v_progr02}
	\end{subfigure}
	$\qquad\qquad$
	\begin{subfigure}[b]{0.23\textwidth}
		\centering
		\includegraphics[width=\textwidth]{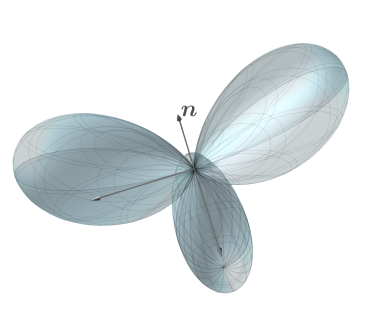}
		\caption{$S=10/3$ and $b=0$}
		\label{fig:c2v_progr03}
	\end{subfigure}
	
	\begin{subfigure}[b]{0.23\textwidth}
		\centering
		\includegraphics[width=\textwidth]{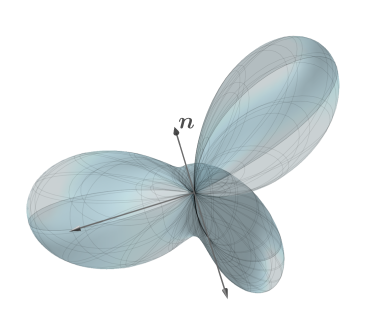}
		\caption{$S=1.67$ and $b=2.31$}
		\label{fig:c2v_progr04}
	\end{subfigure}
	$\qquad\qquad$
	\begin{subfigure}[b]{0.23\textwidth}
		\centering
		\includegraphics[width=\textwidth]{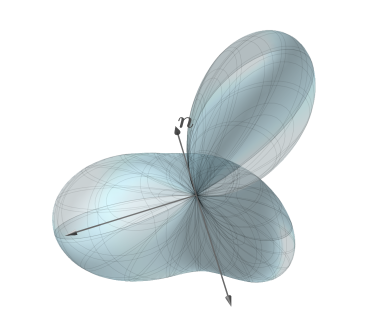}
		\caption{$S=-1.39$ and $b=4.31$}
		\label{fig:c2v_progr05}
	\end{subfigure}
	$\qquad\qquad$
	\begin{subfigure}[b]{0.23\textwidth}
		\centering
		\includegraphics[width=\textwidth]{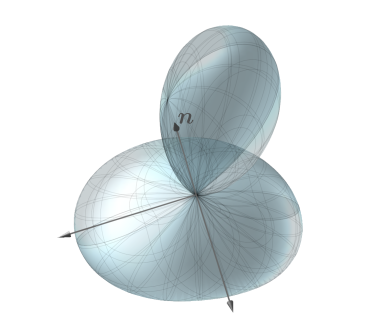}
		\caption{$S=-15.37$ and $b=9.02$}
		\label{fig:c2v_progr06}
	\end{subfigure}
	\caption{Following the blue trajectory on plane $b_{2}=0$: sequence of octupolar potentials with $\mathsf{D}_{2 h}$ symmetry for $\beta=0$. Starting from an almost pure bend (a), the trajectory touches the monkey saddle (c) and then tends to the pure splay with $\mathsf{D}_{\infty h}$ symmetry as $b\to+\infty$ (and $S\to-\infty$).}
	\label{fig:c2v_progression}
\end{figure}

\begin{figure}[h]
	\centering
	\begin{subfigure}[b]{0.23\textwidth}
		\centering
		\includegraphics[width=\textwidth]{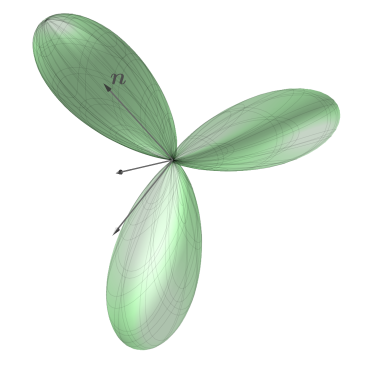}
		\caption{$S=-10/3$ and $b=0$}
		\label{fig:c3v_progr01}
	\end{subfigure}
	$\qquad\qquad$
	\begin{subfigure}[b]{0.23\textwidth}
		\centering
		\includegraphics[width=\textwidth]{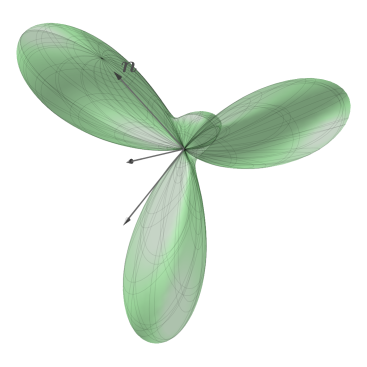}
		\caption{$S=-1.68$ and $b=-0.72$}
		\label{fig:c3v_progr02}
	\end{subfigure}
	$\qquad\qquad$
	\begin{subfigure}[b]{0.23\textwidth}
		\centering
		\includegraphics[width=\textwidth]{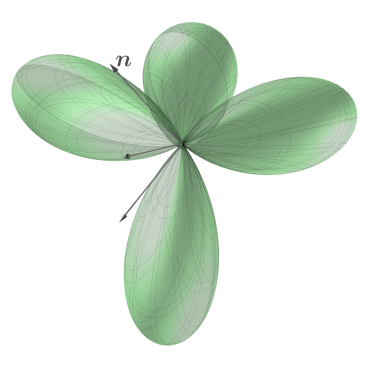}
		\caption{$S=0$ and $b=0$}
		\label{fig:c3v_progr03}
	\end{subfigure}
	
	\begin{subfigure}[b]{0.23\textwidth}
		\centering
		\includegraphics[width=\textwidth]{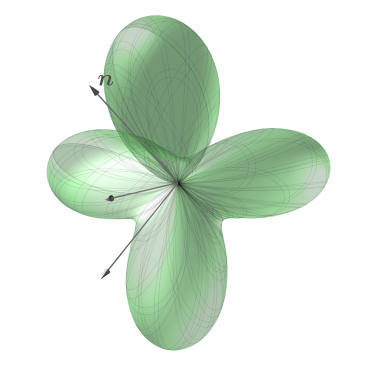}
		\caption{$S=0.07$ and $b=0.97$}
		\label{fig:c3v_progr04}
	\end{subfigure}
	$\qquad\qquad$
	\begin{subfigure}[b]{0.23\textwidth}
		\centering
		\includegraphics[width=\textwidth]{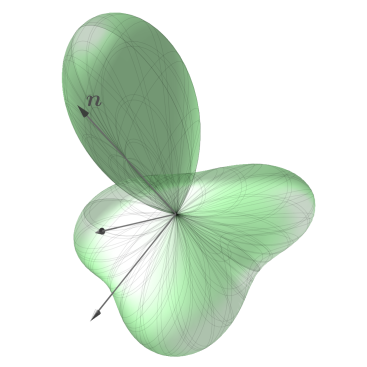}
		\caption{$S=-1.32$ and $b=2.37$}
		\label{fig:c3v_progr05}
	\end{subfigure}
	$\qquad\qquad$
	\begin{subfigure}[b]{0.23\textwidth}
		\centering
		\includegraphics[width=\textwidth]{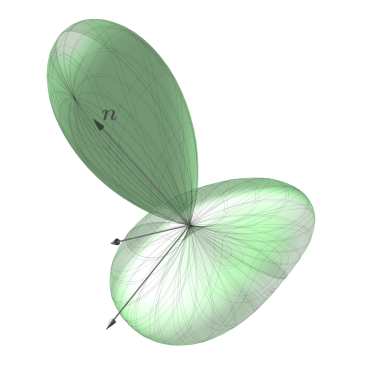}
		\caption{$S=-7.22$ and $b=4.85$}
		\label{fig:c3v_progr06}
	\end{subfigure}
	\caption{Following the green trajectory on plane $b_{2}=0$: sequence of  octupolar potentials with $\mathsf{D}_{3d}$ symmetry for $\beta=0$. Starting from the monkey saddle (a), the trajectory touches the tetrahedron (c) and then tends to the pure splay with $\mathsf{D}_{\infty h}$ symmetry as $b$ increases.}
	\label{fig:c3v_progression}
\end{figure}

\paragraph{$\mathsf{D}_{6h}$.} A very special situation occurs for $b=0$ and  $S=\pm 10/3$. Here the octupolar potential has three equal lobes spatially distributed as the vertices of an equilateral triangle.  The symmetry group is $\mathsf{D}_{6h}$, with rotation axes (by angle $\pi/3$) given by $\vt$ (for $S$ positive) and $\vo$ (for $S$ negative); see Fig.~\ref{fig:monkey_saddle}.
It is yet another manifestation of  the \emph{monkey saddle} defined in \cite[p.\,191]{hilbert:geometry}.\footnote{In its original realization, the monkey saddle is a surface in $3\mathrm{D}$ with three depressions, instead of the two needed for a human rider (the third accommodating the monkey's tail).}  The corresponding  points in the space of distortion characteristics  $(S,b,\beta)$  are marked in red in Fig.~\ref{fig:symmetries}.
\paragraph{$\mathsf{D}_{2h}$.} On the whole $S$ axis, $\Phi$ enjoys the lower $\mathsf{D}_{2 h}$ symmetry. Other loci with the same symmetry  
are the straight lines $S=0$ and $\beta=\pm\frac{\pi}{4}$ (the bisectrices  $b_{2}=\pm b_{1}$ in Fig.~\ref{fig:sym_b1b2}) together with four extra (blue) curves passing through the monkey-saddle points in Figs.~\ref{fig:sym_sb1} and \ref{fig:sym_sb2}. The latter lie in pair in the planes $\beta=0$ (Fig.~\ref{fig:sym_sb1}) and $\beta=\frac{\pi}{2}$ (Fig.~\ref{fig:sym_sb2}),  bounded on one side  by the horizontal asymptotes $S\doteq\pm4.00$, and increasing (or decreasing) at the other side as $S/b^2\doteq\pm0.22$.
Each curve intersects only once the plane $S=0$, at   $b=\frac{5}{\sqrt{2}}$; there the polar plot of the  octupolar potential has one lobe surmounting two equal legs (Fig.~\ref{fig:c2v}). Figure~\ref{fig:c2v_progression} shows the sequence of potentials along one of the $\mathsf{D}_{2 h}$ trajectories.
\paragraph{$\mathsf{D}_{3h}$.} More dramatic are the loci in space $(S,b,\beta)$ where $\Phi$ exhibits the $\mathsf{D}_{3d}$ symmetry. These are lines, in the planes $\beta=0$ and $\beta=\frac{\pi}{2}$, making a loop between the monkey saddles and the tetrahedron of pure biaxial splay (see the green lines in Figs.~\ref{fig:sym_sb1} and \ref{fig:sym_sb2}). They grow parabolically as $S/b^2\doteq\pm0.33$ and intersect the plane $S=0$ at $b=\frac{\sqrt{5}}{2}$, where the polar plot of the octupolar potential is a tripod surmonted by a single lobe (see Fig.~\ref{fig:c3v}). Figure~\ref{fig:c3v_progression} shows the sequence of potentials along one of these $\mathsf{D}_{3d}$ trajectories.
\paragraph{$\mathsf{C}_{2h}$.}
Finally, we consider the smallest symmetry group, $\mathsf{C}_{2 h}$, which is a subgroup of all the other groups discussed here, so that all the points in the aforementioned trajectories are also invariant under a single reflection.
By combining \eqref{eq:sym1} and \eqref{eq:sym3}, it is straightforward to obtain 
\begin{equation}\label{eq:sym_biaxial}
\Phi(\bm{x};S,b,0,1)=\Phi(\tn{R}_{\vt}\bm{x};S,b,0,1) 
\qquad\text{and}\qquad 
\Phi(\bm{x};S,b,\pi/2,1,\bm{x})=\Phi(\tn{R}_{\vo}\bm{x};S,b,\pi/2,1)\,;
\end{equation} 
therefore both planes $b_{2}=0$ ($\beta=0$) and $b_{1}=0$ ($\beta=\pi/2$) enjoy the invariance under $\mathsf{C}_{2 h}$. For $\beta\neq0$ and $\beta\neq\frac\pi2$, the $\mathsf{C}_{2h}$ symmetry survives only in selected places; they are the fancy purple lines depicted in  Figs.~\ref{fig:sym_Pi32}-\ref{fig:sym_Pi4} for different values of $\beta$. All these lines connect the monkey saddles to the tetrahedron and have an asymptote that drifts from $S=4$ to $S=0$ as $\beta$ increases from $0$ to $\frac{\pi}{4}$. For $\beta=\frac{\pi}{4}$, the asympote abruptly closes down and is promoted to the higher symmetry group $\mathsf{D}_{2h}$. On the other hand, when $\beta$ reaches $0$, the closed $\mathsf{C}_{2h}$ trajectory abruptly splits up in two separate open trajectories, promoted to two distinct higher symmetry groups, namely, $\mathsf{D}_{2h}$ and  $\mathsf{D}_{3d}$.

Figure~\ref{fig:symmetries} shows that blue, green, and purple lines are all symmetric with respect to the  $S$ axis and, in addition, the blue  and green lines flip on opposite sides of the $S$ axis in swapping the planes $b_{2}=0$ and $b_{1}=0$. These features are clear consequences of  properties \eqref{eq:symmetries}, which also explain why the most representative  $\mathsf{C}_{2 h}$ trajectories in Fig.~\ref{fig:symmetries} are those for $\beta\in[0,\frac\pi4]$: all other trajectories are obtained by rotations and reflections of these. For example, the trajectory for $\beta=\frac{\pi}{8}$ is  the same as that  for $\beta=-\frac{\pi}{8}$, and the trajectories  for $\beta=\frac{\pi}{8}$ and $\beta=\frac{3\pi}{8}$ differ from one another by a reflection with respect to the $b$ axis.

\subsection{Separatrix}\label{sec:separatrix}
Besides symmetry, a key feature that distinguishes one octupolar potential from the other for the pure modes in Fig.~\ref{fig:pure_modes} is the number of their maxima. Except for the very special case of pure splay, such a number can be either $3$ or $4$ \cite{gaeta:octupolar,gaeta:symmetries}. The critical points of the octupolar potential are the roots  $\bm{x}\in\mathbb{S}^{2}$ of $\grad\Phi(\bm{x})=3\lambda\bm{x}$; there, $\lambda=\Phi(\bm{x})$. In our study, the problem of finding these points is  equivalent to solve the system
\begin{equation}\label{eq:critical_points}
\left\{
\begin{aligned}
 (3S+10q)x_{1}x_{3} + 2b_{2} x_{1}x_{2} + 3b_{1} x_{1}^{2} + b_{1} x_{2}^{2} - 4b_{1} x_{3}^{2}
 &= 15\lambda x_{1}, \\
 (3S-10q)x_{2}x_{3} + 2b_{1} x_{1}x_{2} + b_{2} x_{1}^{2} + 3b_{2} x_{2}^{2} - 4b_{2} x_{3}^{2}
 &= 15\lambda x_{2}, \\
 (3S+10q)x_{1}^{2} + (3S-10q)x_{2}^{2} - 16b_{1} x_{1}x_{3} - 16b_{2} x_{2}x_{3} - 6Sx_{3}^{2}
 &= 30\lambda x_{3},
\end{aligned}
\right.
\end{equation}
over the constraint $x_{1}^{2}+x_{2}^{2}+x_{3}^{2}=1$. As already noticed, here we cannot freely \emph{orient} the potential in the frame $\basis$, as was done in \cite{gaeta:octupolar,gaeta:symmetries}. There, a special surface in parameter space was identified, called the \emph{separatrix}, as it separates a region with $3$ maxima of $\Phi$ from a region with $4$ maxima. That analysis does not apply verbatim to the present setting (and neither does the explicit, algebraic characterization of the separatrix in \cite{chen:octupolar}). In the present context, the search for the separatrix must start afresh. And fresh is its flavor: on the separatix a new direction of distortion concentration either arises or dies away, depending on which side we look at. 

The case $q=0$, where the biaxial splay vanishes, is fairly simple. The potential has always three maxima and four saddle points, except for the very special case of pure splay, where the absolute maximum is surrounded by a ring of idential local maxima.

\begin{figure}[h]
\centering
 \begin{subfigure}[b]{0.23\textwidth}
 \centering
  \includegraphics[width=\textwidth]{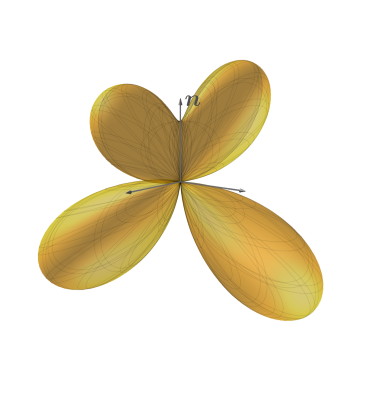}
  \caption{$S=-2/3$}
  \label{fig:progr2}
 \end{subfigure}
 $\ $
 \begin{subfigure}[b]{0.23\textwidth}
 \centering
  \includegraphics[width=\textwidth]{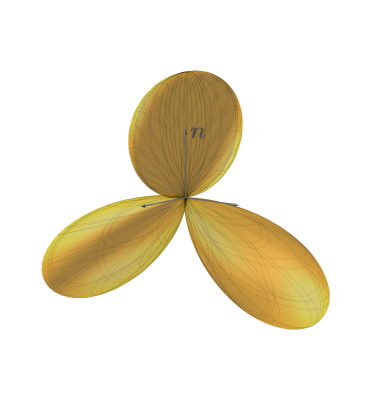}
  \caption{$S=-5/3$}
  \label{fig:progr5}
 \end{subfigure}
 $\ $
 \begin{subfigure}[b]{0.23\textwidth}
 \centering
  \includegraphics[width=\textwidth]{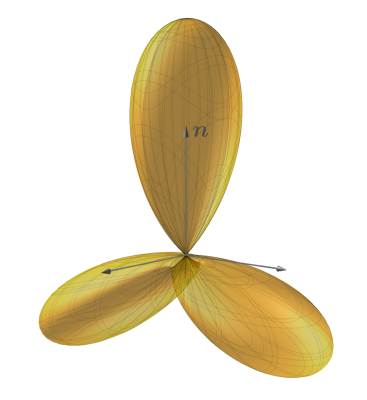}
  \caption{$S=-15/3$}
  \label{fig:progr15}
 \end{subfigure}
 $\ $
 \begin{subfigure}[b]{0.23\textwidth}
 \centering
  \includegraphics[width=\textwidth]{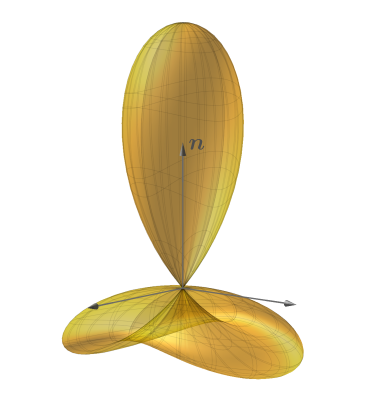}
  \caption{$S=-50/3$}
  \label{fig:progr50}
 \end{subfigure}
\caption{Sequence of octupolar potentials for $q=1$ and $b=0$, as $S\leq0$ decreases.}
\label{fig:s_to_infinity}
\end{figure}

The generic case, where  $q=1$, is more interesting. By a direct inspection of \eqref{eq:critical_points}, we can provide a full description of the critical points when $b=0$ (see Table \ref{tab:stationary} in Appendix A): two of the four maxima of the tetrahedron at $S=0$ merge as $|S|$ increases, until $|S|=\frac{5}{3}$, where they completely coalesce (together with the saddle point between them) in a single maximum. This latter is smaller than the other two maxima at the beginning, but becomes equal to them at $|S|=\frac{10}{3}$, as a  monkey saddle is reached. As $|S|$ increases further, the so far distinct  equal maxima start bridging up, approaching the circular pedestal of the pure splay case, when $S\to\pm\infty$; a representation of this sequence is shown in Fig.~\ref{fig:s_to_infinity}.

\begin{figure}[h]
	\centering
	\begin{subfigure}[b]{0.27\textwidth}
		\centering
		\includegraphics[width=\textwidth]{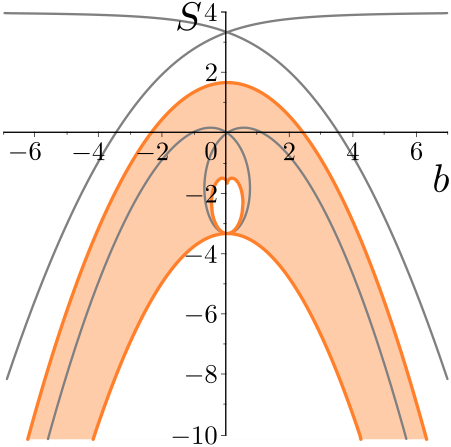}
		\caption{$\beta=0$}
		\label{fig:sep0}
	\end{subfigure}
	$\qquad\qquad$
	\begin{subfigure}[b]{0.27\textwidth}
		\centering
		\includegraphics[width=\textwidth]{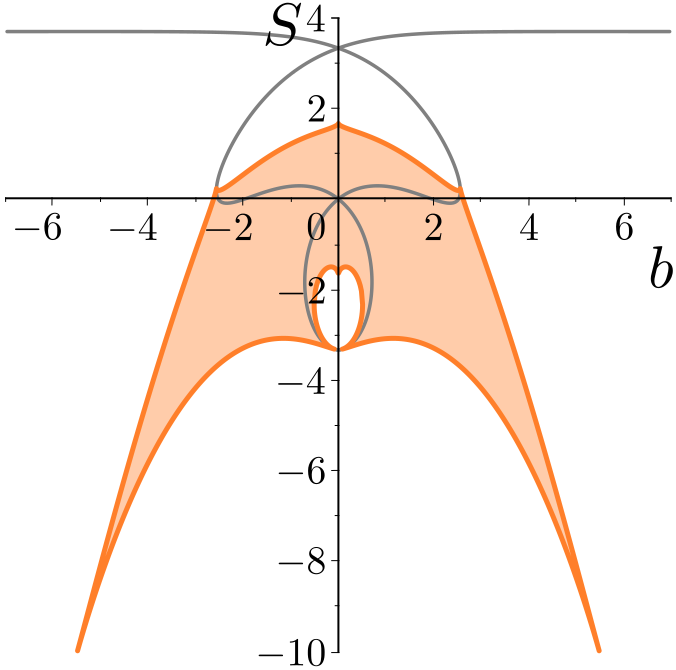}
		\caption{$\beta=\pi/16$}
		\label{fig:sepPi16}
	\end{subfigure}
	
	\vspace{0.5cm}
	
	\begin{subfigure}[b]{0.27\textwidth}
		\centering
		\includegraphics[width=\textwidth]{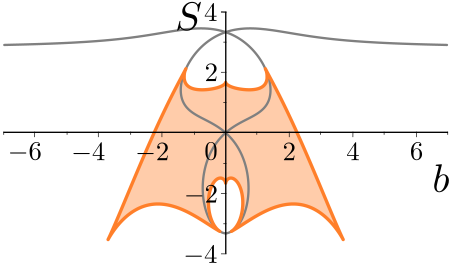}
		\caption{$\beta=\pi/8$}
		\label{fig:sepPi8}
	\end{subfigure}
	$\qquad\qquad$
	\begin{subfigure}[b]{0.27\textwidth}
		\centering
		\includegraphics[width=\textwidth]{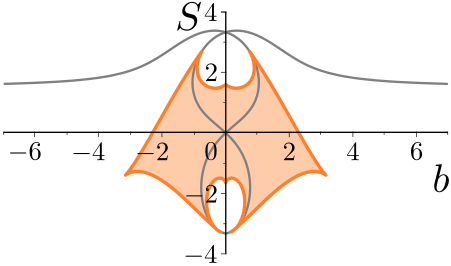}
		\caption{$\beta=3\pi/16$}
		\label{fig:sep3Pi16}
	\end{subfigure}
	$\qquad\qquad$
	\begin{subfigure}[b]{0.27\textwidth}
		\centering
		\includegraphics[width=\textwidth]{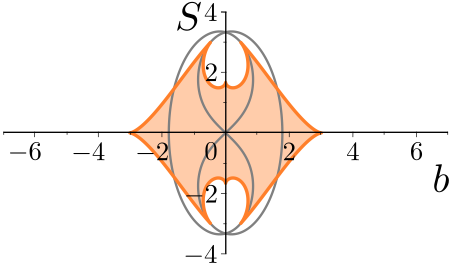}
		\caption{$\beta=\pi/4$}
		\label{fig:sepPi4}
	\end{subfigure}
	\caption{Profiles of the separatrix for different $\beta\in[0,\frac{\pi}{4}]$ and the conjugated values $\beta+\pi$, represented by $b\leq0$, as in Fig.~\ref{fig:symmetries}. The regions with four maxima are in orange; all include the origin. The gray lines are the same symmetry trajectories from Fig.~\ref{fig:symmetries}.}
	\label{fig:separatrix}
\end{figure}

\begin{figure}[h]
	\centering
	\begin{subfigure}[b]{0.23\textwidth}
		\centering
		\includegraphics[width=\textwidth]{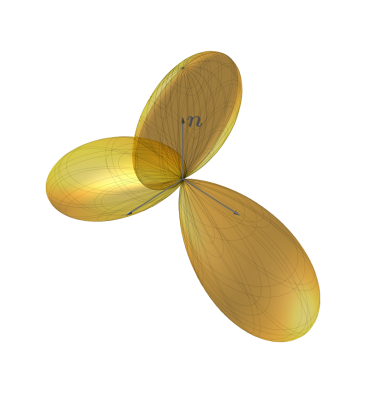}
		\caption{$b=0.1$}
		\label{fig:sep_progression01}
	\end{subfigure}
	$\ $
	\begin{subfigure}[b]{0.23\textwidth}
		\centering
		\includegraphics[width=\textwidth]{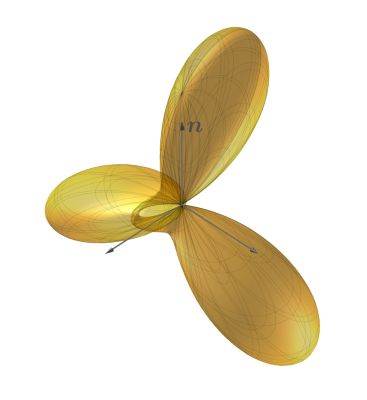}
		\caption{$b=0.7$}
		\label{fig:sep_progression07}
	\end{subfigure}
	$\ $
	\begin{subfigure}[b]{0.23\textwidth}
		\centering
		\includegraphics[width=\textwidth]{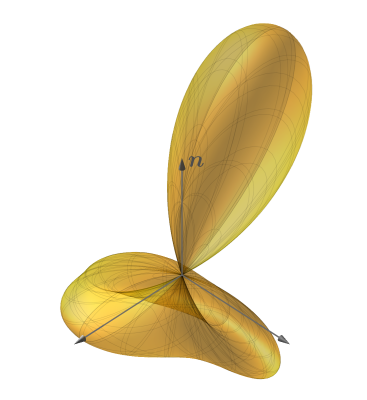}
		\caption{$b=3$}
		\label{fig:sep_progression3}
	\end{subfigure}
	$\ $\begin{subfigure}[b]{0.23\textwidth}
		\centering
		\includegraphics[width=\textwidth]{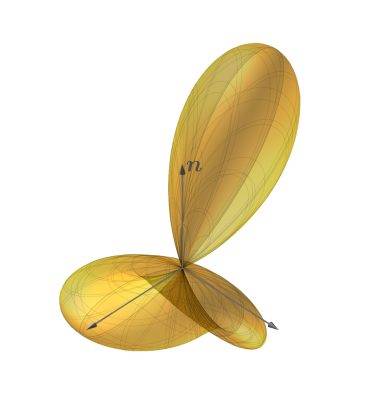}
		\caption{$b=5$}
		\label{fig:sep_progression5}
	\end{subfigure}
	\caption{Sequence of octupolar potentials for $q=1$, $S=-2$ and $\beta=0$, for increasing values of $b$. At the beginning the maxima are $3$; after the first crossing of the separatrix they become $4$ and then,  after the second crossing, they are $3$ again.}
	\label{fig:sep_progression}
\end{figure}

It is clear from this example that  the very special points $|S|=\frac{5}{3}$, where two maxima and one saddle of $\Phi$ coalesce into a single maxima, belong to the separatrix. These points also separate on  the $S$ axis  an inner interval ($|S|<\frac53$), where the potential resembles a pure biaxial splay, from two outer intervals ($|S|>\frac53$), where the octupolar potential resembles a pure splay. Thus, the octupolar potential reveals the presence of biaxial splay in a generic distortion: when $\Phi$ has four maxima, the biaxial splay is, in general, predominant with respect to the other modes, whereas when $\Phi$ has three maxima there is little or no biaxial splay. More details are offered by the profiles of the separatrix depicted in Fig.~\ref{fig:separatrix}; they were obtained by numerical continuation of the new maximum arising at $|S|=\frac{5}{3}$.
The region with four maxima encloses the origin (where the pure biaxial splay sits), whereas the region with three maxima spreads  away from the origin. Only for $\beta=0$, the four-maxima region is not bounded and extends in territories with large  $|S|$ and $b$, but in these cases the presence of an almost circular pedestal in the potential witnesses a net dominance of the splay mode. As shown in Fig.~\ref{fig:separatrix}, the four-maxima region enclosed by the separatrix is not convex. This causes a sort of ``re-entrant'' effect illustrated in Fig.~\ref{fig:sep_progression}.
For a given (appropriately chosen) $S$, upon increasing $b$ from naught, the octupolar potential starts by having exceptionally $3$ maxima; they soon become $4$, which is the expected number, and then, for yet larger values of $b$, they get back to the normal $3$.

In studying the separatrix we also benefited from the symmetry properties \eqref{eq:symmetries}: they again ensure that the sector $\beta\in]0,\frac{\pi}{4}]$ in the space $(S,b,\beta)$ suffices to describe completely the behaviour of the separatrix; all remaining sectors can be obtained by rotations and reflections.

\section{Quasi-Uniform Distortions}\label{sec:defects}
Building upon the decomposition of $\gradn$ in \eqref{eq:grad_n}, it is natural to define a \emph{uniform} nematic distortion as one for which all distortion characteristics $(S,T,b_1,b_2,q)$ are constant in space \cite{virga:uniform}. Inside a uniform distortion, it is as if, sitting in a place in space, one sees the same orientation landscape against the local distortion frame $\dframe$ (which is intrinsically defined), irrespective of the place. It was shown in \cite{virga:uniform} that the family of all uniform distortions is characterized by having $S=0$ and $T=\pm2q$ with, correspondingly, $b_1=\pm b_2$. Moreover, this  two-fold family is exhausted by the \emph{heliconical} director fields first envisioned by Meyer \cite{meyer:structural} and which have recently been identified experimentally with the ground state of \emph{twist-bend} nematic phases \cite{cestari:phase}.\footnote{A vast literature has lately grown on twist-bend phases and their still intriguing germination out of the traditional nematic phase. On the theoretical side, a fair representation of the variety of available contributions is offered by the papers \cite{dozov:spontaneous,shamid:statistical,virga:double-well,kats:landau,greco:molecular,barbero:elastic,tomczyk:twist-bend,osipov:effect,vanakaras:molecular,lelidis:nonlinear,barbero:fourth-order,aliev:helicoidal}. On the experimental side, the following papers are among the most relevant, \cite{chen:chiral,chen:twist-bend,borshch:nematic,gorecka:short,paterson:understanding,salamonczyk:structure,tuchband:distinct,zhu:resonant}.}

\begin{figure}[h]
	\centering
	\begin{subfigure}[b]{0.23\textwidth}
		\centering
		\includegraphics[width=\textwidth]{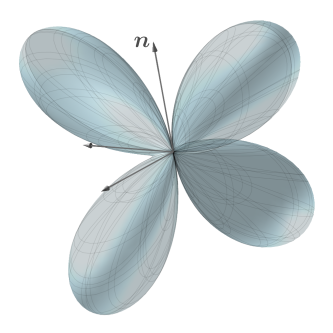}
		\caption{$b=0.2$}
		\label{fig:twist02}
	\end{subfigure}
	$\ $
	\begin{subfigure}[b]{0.23\textwidth}
		\centering
		\includegraphics[width=\textwidth]{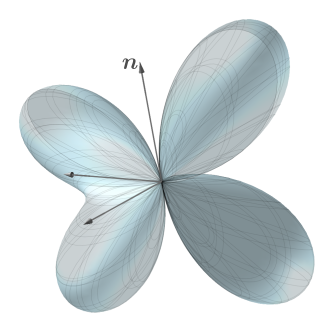}
		\caption{$b=0.8$}
		\label{fig:twist08}
	\end{subfigure}
	$\ $
	\begin{subfigure}[b]{0.23\textwidth}
		\centering
		\includegraphics[width=\textwidth]{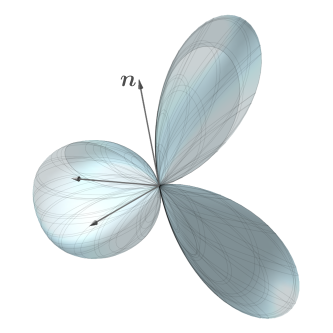}
		\caption{$b=2.5$}
		\label{fig:twist25}
	\end{subfigure}
	$\ $
	\begin{subfigure}[b]{0.23\textwidth}
		\centering
		\includegraphics[width=\textwidth]{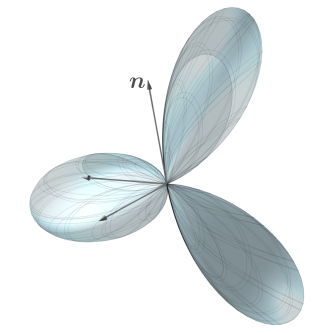}
		\caption{$b=5.0$}
		\label{fig:twist50}
	\end{subfigure}
	\caption{Sequence of octupolar potentials for $q=1$, $S=0$, and $\beta=\pi/4$, with increasing $b\geq0$. These pictures are in accord with Fig.~\ref{fig:sepPi4}; the separatrix is crossed at $b\doteq3.02$.}
	\label{fig:twist_bend}
\end{figure}

Our ``octupolar eye'' does not see $T$ and, as already remarked, it is also insensitive to rescaling (by a constant) all distortion characteristics.\footnote{This is why above we could take $q=1$ with no loss of generality.} The octupolar potential (and its graphical representation) is thus especially suited to describe uniform distortions. In the notation used in this paper, the latter correspond to $q=1$ (and $T=\pm2$), $S=0$, $\beta=\pm\frac\pi4$, and arbitrary $b\geq0$. As shown in Fig.~\ref{fig:sym_b1b2}, for $b>0$ the octupolar potential of uniform distortions enjoys the $\mathsf{D}_{2h}$ symmetry (see Fig.~\ref{fig:twist_bend}).

It is precisely this way of representing uniform distortions to suggest the definition of a larger class of distortions: those with (spatially) uniform (rescaled) octupolar potentials. This is the case whenever the (non-vanishing) distortion characteristics are not necessarily constant themselves, but are in a constant ratio to one another, thus ensuring that the octupolar potential (with its directions of distortion concentration) has the same appearance everywhere in space. We shall call these distortions \emph{quasi-uniform}. They are formally defined in terms of the distortion characteristics as follows: either $(S,T,b_1,b_2,q)$ are all zero but one, or those that do not vanish are all proportional through constants to a certain non-vanishing function of position. Whenever possible, as above, we shall conventionally rescale all distortion characteristics to $q$. In words, a quasi-uniform distortion is a uniform distortion rescaled differently in different positions in space.

\subsection{Examples}\label{sec:examples}
\begin{figure}[h]
	\centering
	\begin{subfigure}[b]{0.19\textwidth}
		\centering
		\includegraphics[width=0.95\textwidth]{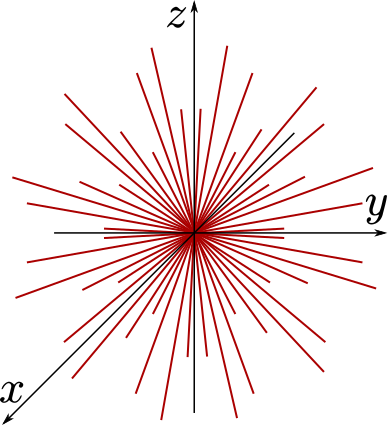}
		\caption{Hedgehog}
		\label{fig:hedgehog}
	\end{subfigure}
	\begin{subfigure}[b]{0.19\textwidth}
		\centering
		\includegraphics[width=0.95\textwidth]{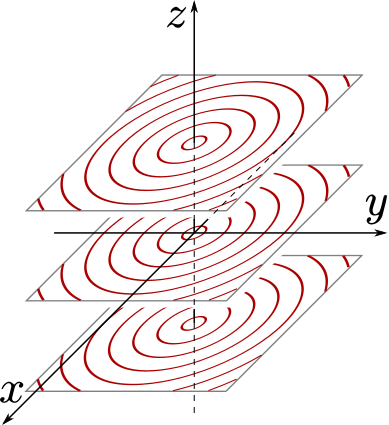}
		\caption{Pure bend}
		\label{fig:concentric}
	\end{subfigure}
	\begin{subfigure}[b]{0.19\textwidth}
		\centering
		\includegraphics[width=0.95\textwidth]{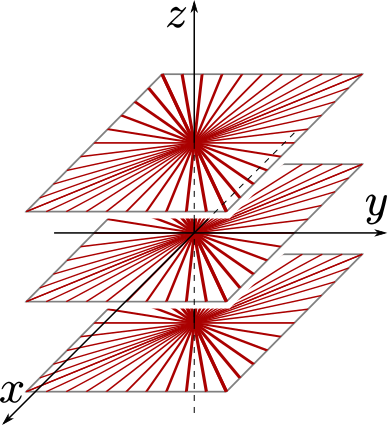}
		\caption{Planar splay}
		\label{fig:planar}
	\end{subfigure}
	\begin{subfigure}[b]{0.19\textwidth}
		\centering
		\includegraphics[width=0.95\textwidth]{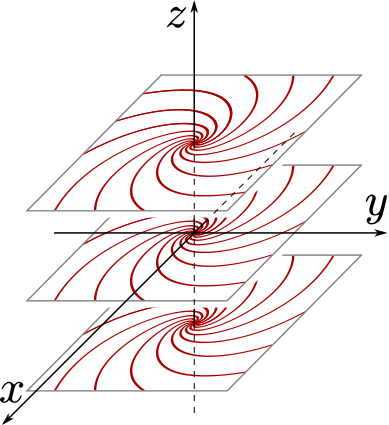}
		\caption{Spirals ($\alpha=\pi/4$)}
		\label{fig:parabolic}
	\end{subfigure}
	\begin{subfigure}[b]{0.19\textwidth}
		\centering
		\includegraphics[width=0.95\textwidth]{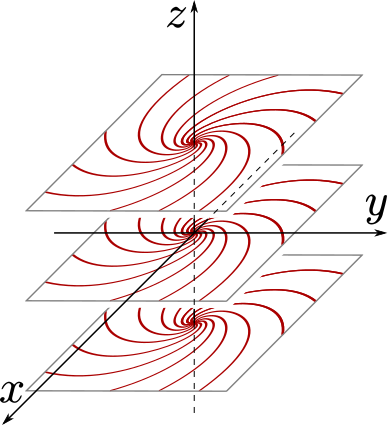}
		\caption{Spirals ($\alpha=-\pi/4$)}
		\label{fig:spiral}
	\end{subfigure}
	\caption{Integral lines for examples of quasi-uniform distortions.}
	\label{fig:defects}
\end{figure}

\begin{figure}[h]
	\centering
	\begin{subfigure}[b]{0.19\textwidth}
		\centering
		\includegraphics[width=0.95\textwidth]{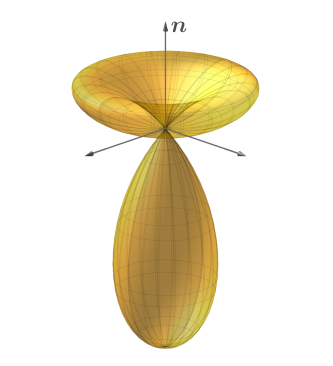}
		\caption{Hedgehog}
		\label{fig:pot_hedgehog}
	\end{subfigure}
	\begin{subfigure}[b]{0.19\textwidth}
		\centering
		\includegraphics[width=0.95\textwidth]{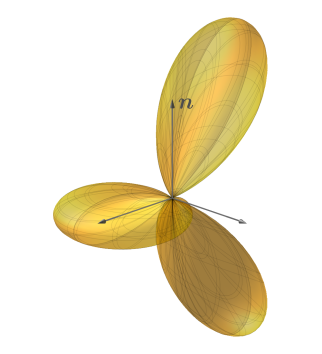}
		\caption{Pure bend}
		\label{fig:pot_concentric}
	\end{subfigure}
	\begin{subfigure}[b]{0.19\textwidth}
		\centering
		\includegraphics[width=0.95\textwidth]{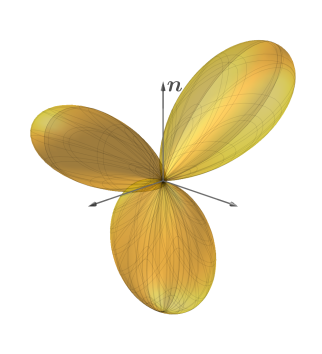}
		\caption{Planar splay}
		\label{fig:pot_planar}
	\end{subfigure}
	\begin{subfigure}[b]{0.19\textwidth}
		\centering
		\includegraphics[width=0.95\textwidth]{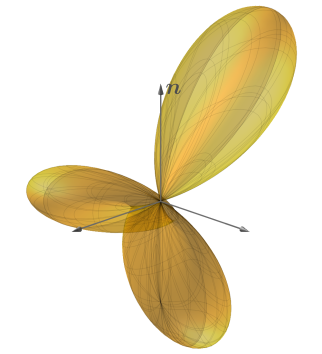}
		\caption{Spirals ($\alpha=\pi/4$)}
		\label{fig:pot_parabolic}
	\end{subfigure}
	\begin{subfigure}[b]{0.19\textwidth}
		\centering
		\includegraphics[width=0.95\textwidth]{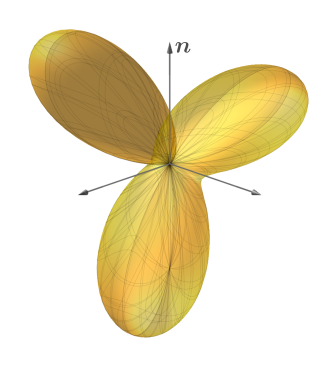}
		\caption{Spirals ($\alpha=-\pi/4$)}
		\label{fig:pot_spiral}
	\end{subfigure}
	\caption{Octupolar potentials for the quasi-uniform distortions in Fig.~\ref{fig:defects}.}
	\label{fig:pot_defects}
\end{figure}

Quasi-uniform distortions have not yet been fully characterized. For illustrative purposes, we have worked out a number of simple examples, which are briefly summarized below.
Here we shall employ a given Cartesian frame $\xyz$ with origin $o$ in which the position vector  of a generic point $p$ is $p-o=x\vx+y\vy+z\vz$.  The nematic director field $\vn$ will be described by its components in the frame $\xyz$ expressed as functions of  the coordinates $(x,y,z)$. The integral lines of the examples presented in this section are shown in Fig.~\ref{fig:defects}.
The corresponding octupolar potentials have polar plots shown in Fig.~\ref{fig:pot_defects}.

\subsubsection{Hedgehog}
The first example we consider is the \emph{hedgehog},
\begin{equation}\label{eq:hedgehog}
 \vn= \frac{1}{\sqrt{x^{2}+y^{2}+z^{2}}}\left(x\vx+y\vy+z\vz\right),
\end{equation}
 which results in a pure splay with 
 \begin{equation}\label{eq:hedgehog_char}
 S=\frac{2}{\sqrt{x^{2}+y^{2}+z^{2}}},\quad
 T=b=q=0.
 \end{equation}
 Scaling the distortion characteristics to $S$, as in Sec.~\ref{sec:vanishing}, we obtain the octupolar potential $\Phi(\bm{x};1,0,-,0)$, which enjoys the   $\mathsf{D}_{\infty h}$ symmetry (see Fig.~\ref{fig:pot_hedgehog}).

\subsubsection{Pure bend}
A pure bend is represented by the director field
\begin{equation}\label{eq:circles}
 \vn= \frac{1}{\sqrt{x^{2}+y^{2}}} \left(-y\vx+x\vy\right),
\end{equation}
whose integral lines are concentric circles in the planes with constant $z$. In this case,
\begin{equation}\label{eq:pure_bend}
S=T=q=0,\quad \bend=\frac{1}{x^{2}+y^{2}} \left(x\vx+y\vy\right).
\end{equation}
Scaled to $b$, the
octupolar potential becomes $\Phi(\bm{x};0,1,0,0)$, which enjoys the $\mathsf{D}_{2h}$ symmetry (see Fig.~\ref{fig:pot_concentric}).

\subsubsection{Planar splay}
A pure splay in two space dimensions is described by the field
\begin{equation}\label{eq:planar}
 \vn= \frac{1}{\sqrt{x^{2}+y^{2}}} \left(x\vx+y\vy\right),
\end{equation}
which we call a \emph{planar splay}. In this case,
\begin{equation}\label{eq:planar_splay}
S=\frac{1}{\sqrt{x^{2}+y^{2}}},\quad T=b=0,\quad q=\frac{1}{2\sqrt{x^{2}+y^{2}}}.
\end{equation}
Scaled to $q$, the corresponding octupolar potential becomes $\Phi(\bm{x};2,0,-,1)$ (see Fig.~\ref{fig:pot_planar}); as already seen in Sec.~\ref{sec:pure_modes}, it differs from the potential of pure bend only in its spatial orientation with respect to $\vn$.
\subsubsection{Spirals}
A somewhat intermediate distortion between pure bend and planar splay is 
given by
\begin{equation}\label{eq:spirals}
 \vn= \frac{1}{\sqrt{x^{2}+y^{2}}} \left[(x\cos\alpha-y\sin\alpha)\vx+(x\sin\alpha+y\cos\alpha)\vy\right],\quad 0<\alpha<\frac\pi2.
\end{equation}
For $\alpha=0$, this field is the pure bend \eqref{eq:pure_bend}, whereas for $\alpha=\frac\pi2$ it is the planar splay \eqref{eq:planar_splay}.  For $0<\alpha<\frac\pi2$, the integral lines of \eqref{eq:spirals} are \emph{spirals} in the planes with constant $z$ (see Figs.~\ref{fig:parabolic}-\ref{fig:spiral}). Scaled to $q$, the octupolar potential associated with \eqref{eq:spirals} becomes $\Phi(\bm{x};2,2\tan\alpha,0,1)$ (see Figs.~\ref{fig:pot_parabolic}-\ref{fig:pot_spiral}). In this case,
\begin{equation}\label{eq:spirals_characteristics}
S = \frac{\cos\alpha}{\sqrt{x^{2}+y^{2}}},\quad T=0,\quad \bend = \frac{\sin\alpha}{\sqrt{x^{2}+y^{2}}}\vo,\quad q =\frac{\cos\alpha}{2\sqrt{x^{2}+y^{2}}},
\end{equation}
where
\begin{equation}\label{eq:spirals_n_1_n_2}
\vo = \frac{1}{\sqrt{x^{2}+y^{2}}} \left[
(x\sin\alpha+y\cos\alpha)\vx
- (x\cos\alpha-y\sin\alpha)\vy
\right]
\quad\text{and} \quad
\vt = -\vz.
\end{equation}

\section{Conclusions}\label{sec:conclusion}
Describing the distortion in space of a nematic director field $\n$ is easier said than done. Motivated by a fresh look \cite{selinger:interpretation} into the classical topic of nematic elasticity, which originated with the works of Oseen~\cite{oseen:theory} and Frank~\cite{frank:theory}, we devised a mathematical construct that can easily identify various independent elastic modes, especially the biaxial splay, which is contending the role traditionally played in liquid crystal science by saddle-splay elasticity \cite{selinger:interpretation}. 

Our construct is an octupolar tensor $\oct$, that is, a third-rank, fully symmetric and traceless tensor built from $\n$ and $\gradn$. More precisely, the octupolar potential $\Phi$ associated with $\oct$ and its graphical representation over the unit sphere $\sphere$ (the polar plot) were particularly expedient in representing the directions of local distortion concentration, which were identified with the directions along which $\Phi$ attains a local maximum. We charted the symmetries enjoyed by $\Phi$ (and its polar graph) in an intrinsic distortion frame (which includes $\n$), as some distortion characteristics of the nematic field are varied in a three-dimensional parameter space. 

Different distortion landscapes correspond to different octupolar potentials, differently oriented relative to $\n$. It is our hope that the symmetries charted in Fig.~\ref{fig:symmetries} could guide the educated eye to recognize which elastic mode is prevailing in a generic distortion. The least we did was to establish a rule of thumb. In general, the octupolar potential $\Phi$ at a point in space may have either $4$ or $3$ directions of distortion concentration; the former spatial arrangement signals the predominance of biaxial splay, whereas the latter signals its depression. 

In general, $\Phi$ changes from point to point. In one case, it is everywhere the same, that is, when all distortion characteristics are constant in space, as is the case for the uniform distortions, which have been characterized in \cite{virga:uniform}. If one allows $\Phi$ to be differently inflated at different places in space, while keeping everywhere the same shape, one defines a new class of quasi-uniform distortions. This class has not yet been characterized, but it has been illustrated here by example.  

\begin{acknowledgments}
The work of A.P. was supported financially by the Department of Mathematics of the University of Pavia as part of the activities funded by the Italian MIUR under the nationwide Program ``Dipartimenti di Eccellenza (2018-2022)''.  
\end{acknowledgments}

\appendix
\section{Mathematical Details}\label{sec:appendix}
For the interested reader, we collect in this Appendix the mathematical details of our development and the (simple) proofs that would have hampered our presentation.  
\subsection{Critical points of $\Phi$  in a special case}
In table \ref{tab:stationary} we list the critical points $\x\in\sphere$ and the critical values $\Phi(\x)$ of the octupolar potential (i.e., the eigenvalues $\lambda$ of the octupolar tensor $\oct$) in the special case considered in Sec.~\ref{sec:separatrix}, namely, for $b=0$ and $q=1$. 
\begin{table}
\renewcommand{\arraystretch}{3.0}
 \centering
 \footnotesize
 \begin{tabular}{|l|l|l|}
 \hline
 & maxima & saddle points \\
 \hline
 $0\geq S>-\frac{5}{3}$ & $\lambda=\frac{\sqrt{(10-3S)^{3}}}{15\sqrt{15(2-S)}}$, \quad$\x=\frac{\pm2\sqrt{5-3S}\vt - \sqrt{10-3S}\vn}{\sqrt{15(2-S)}}$ & $\lambda=-\frac{S}{5}$,\quad $\x=\vn$ \\
 & $\lambda=\frac{\sqrt{(10+3S)^{3}}}{15\sqrt{15(2+S)}}$,\quad  $\x=\frac{\pm2\sqrt{5+3S}\vo + \sqrt{10+3S}\vn}{\sqrt{15(2+S)}}$ & $\lambda=0$,\quad $\x=\frac{\pm\sqrt{10-3S}\vo \pm \sqrt{10+3S}\vt}{2\sqrt{5}}$ \\
 & & $\lambda=\frac{S}{5}$,\quad  $\x=-\vn$ \\[10pt]
 \hline
 $-\frac{5}{3}\geq S>-\frac{10}{3}$,\quad & $\lambda=-\frac{S}{5}$,\quad  $\x=\vn$ & $\lambda=0$,\quad  $\x=\frac{\pm\sqrt{10-3S}\vo \pm \sqrt{10+3S}\vt}{2\sqrt{5}}$ \\
 & $\lambda=\frac{\sqrt{(10-3S)^{3}}}{15\sqrt{15(2-S)}}$,\quad  $\x=\frac{\pm2\sqrt{5-3S}\vt - \sqrt{10-3S}\vn}{\sqrt{15(2-S)}}$  & \\[10pt]
 \hline
 $-\frac{10}{3}>S$ & $\lambda=-\frac{S}{5}$,\quad  $\x=\vn$ & $\lambda=\frac{\sqrt{-(10+3S)^{3}}}{15\sqrt{-15(2+S)}}$,\quad  $\x=\frac{\pm2\sqrt{-5-3S}\vo - \sqrt{-10-3S}\vn}{\sqrt{-15(2+S)}}$ \\
 & $\lambda=\frac{\sqrt{(10-3S)^{3}}}{15\sqrt{15(2-S)}}$,\quad  $\x=\frac{\pm2\sqrt{5-3S}\vt - \sqrt{10-3S}\vn}{\sqrt{15(2-S)}}$ & $\lambda=-\frac{\sqrt{-(10+3S)^{3}}}{15\sqrt{-15(2+S)}}$,\quad  $\x=\frac{\pm2\sqrt{-5-3S}\vo + \sqrt{-10-3S}\vn}{\sqrt{-15(2+S)}}$\\[10pt]
 \hline
 \end{tabular}
 \normalsize
 \caption{Maxima and saddle points of the octupolar potential for $q=1$, $b=0$,  and $S\leq0$. The minima are antipoldal to the maxima in $\sphere$ and have opposite eigenvalues. The case with $S\geq0$ is similar and can be readily derived from the case $S\leq0$ by use of \eqref{eq:symmetries}. For $S>-5/3$ the critical points are $14$: $4$ maxima, $6$ saddles and $4$ minima. For $S<-5/3$ the critical points point are $10$: $3$ maxima, $4$ saddles and $3$ minima, with the only exception of the monkey saddle at $S=-10/3$, where the saddles are only $2$ (for a total of $8$ critical points).}
 \label{tab:stationary}
\end{table}

\subsection{Symmetry properties}
Equations \eqref{eq:symmetries} are easily  checked, once the actions of a number of relevant symmetries and rotations on  $\bm{x}:=x_{1}\vo+x_{2}\vt+x_{3}\vn$ are duly recorded,
\begin{equation}
\begin{gathered}
 \tn{R}_{\vo}\bm{x} = -x_{1}\vo+x_{2}\vt+x_{3}\vn,\quad
 \tn{R}_{\vt}\bm{x} = x_{1}\vo-x_{2}\vt+x_{3}\vn,\quad
 \tn{R}_{\vn}\bm{x} = x_{1}\vo+x_{2}\vt-x_{3}\vn,\\
 \tn{Q}_{\frac{\pi}{2},\vn}\bm{x} = -x_{2}\vo+x_{1}\vt+x_{3}\vn,\quad
 \tn{R}_{\vt}\tn{Q}_{\frac{\pi}{2},\vn}\bm{x} = -x_{2}\vo-x_{1}\vt+x_{3}\vn.
\end{gathered}
\end{equation}
Then, by direct inspection, we see that 
\begin{eqnarray}
 \Phi(\bm{x};S,b,-\beta,q) &=& \left(\frac{S}{2}+q\right)x_{1}^{2}x_{3} - b\cos\beta x_{1}x_{3}^{2} + b\sin\beta x_{2}x_{3}^{2} + \left(\frac{S}{2}-q\right)x_{2}^{2}x_{3}\nonumber \\
 &
 +& \frac{1}{5}\left(x_{1}^{2}+x_{2}^{2}+x_{3}^{2}\right)\left(b\cos\beta x_{1}-b\sin\beta x_{2}-Sx_{3}\right) = \Phi(\tn{R}_{\vt}\bm{x};S,b,\beta,q),
 \\
 \Phi(\bm{x};-S,b,\pi/2-\beta,q) &=& \left(-\frac{S}{2}+q\right)x_{1}^{2}x_{3} - b\sin\beta x_{1}x_{3}^{2} - b\cos\beta x_{2}x_{3}^{2} + \left(-\frac{S}{2}-q\right)x_{2}^{2}x_{3} \nonumber\\
 &
 +& \frac{1}{5}\left(x_{1}^{2}+x_{2}^{2}+x_{3}^{2}\right)\left(b\sin\beta x_{1}+b\cos\beta x_{2}+Sx_{3}\right) = -\Phi(\tn{R}_{\vt}\tn{Q}_{\frac{\pi}{2},\vn}\bm{x};S,b,\beta,q),
 \\
 \Phi(\bm{x};S,b,\pi+\beta,q) &=& \left(\frac{S}{2}+q\right)x_{1}^{2}x_{3} + b\cos\beta x_{1}x_{3}^{2} + b\sin\beta x_{2}x_{3}^{2} + \left(\frac{S}{2}-q\right)x_{2}^{2}x_{3} \nonumber\\
 &
 +& \frac{1}{5}\left(x_{1}^{2}+x_{2}^{2}+x_{3}^{2}\right)\left(-b\cos\beta x_{1}-b\sin\beta x_{2}-Sx_{3}\right) = -\Phi(\tn{R}_{\vn}\bm{x};S,b,\beta,q).
\end{eqnarray}

Similarly, we prove \eqref{eq:sym_biaxial}. That 
$\Phi(\bm{x};S,b,0,1)=\Phi(\tn{R}_{\vt}\bm{x};S,b,0,1)$ is a trivial consequence of \eqref{eq:sym1}. Then, by
\eqref{eq:sym3}, \eqref{eq:sym1}, and the identity  $\tn{R}_{\vt}\tn{R}_{\vn}\bm{x} = x_{1}\vo-x_{2}\vt-x_{3}\vn = -\tn{R}_{\vo}\bm{x}$, we arrive at
\begin{equation}
\begin{split}
\Phi(\bm{x};S,b,\pi/2,1)&=\Phi(\bm{x};S,b,\pi-\pi/2,1)=-\Phi(\tn{R}_{\vn}\bm{x};S,b,-\pi/2,1)\\
&=-\Phi(\tn{R}_{\vt}\tn{R}_{\vn}\bm{x};S,b,\pi/2,1)=-\Phi(-\tn{R}_{\vo}\bm{x};S,b,\pi/2,1,)=\Phi(\tn{R}_{\vo}\bm{x};S,b,\pi/2,1).
\end{split}
\end{equation}

\end{document}